\documentclass[conference]{IEEEtran}
\IEEEoverridecommandlockouts
\usepackage{amsmath,amssymb,amsfonts}
\usepackage{algorithm}
\usepackage{algorithmicx}
\usepackage{algpseudocode}
\algnewcommand\algorithmicforeach{\textbf{for each}}
\algdef{S}[FOR]{ForEach}[1]{\algorithmicforeach\ #1\ \algorithmicdo}
\algblock{Input}{EndInput}
\algnotext{EndInput}
\algblock{Output}{EndOutput}
\algnotext{EndOutput}
\newcommand{\Desc}[2]{\State \makebox[2em][l]{#1}#2}
\usepackage{graphicx}
\usepackage{caption}
\usepackage{subcaption}
\usepackage{textcomp}
\usepackage{xcolor}
\usepackage{hyperref}
\hypersetup{colorlinks=true,linkcolor=black,citecolor=blue,filecolor=black,urlcolor=blue}
\usepackage{xspace}
\usepackage{multirow}
\usepackage{textcomp}
\usepackage[binary-units]{siunitx}
\def\BibTeX{{\rm B\kern-.05em{\sc i\kern-.025em b}\kern-.08em
    T\kern-.1667em\lower.7ex\hbox{E}\kern-.125emX}}

\newcommand{\bigbrain}{BigBrain\xspace}

\begin{document}

\title{Performance benefits of Intel\textsuperscript{\textregistered}~Optane\texttrademark~DC persistent memory for the parallel processing of large neuroimaging data}

\author{\IEEEauthorblockN{Val\'erie Hayot-Sasson$^1$, Shawn T Brown$^2$ and 
    Tristan Glatard$^1$
  }\\
  \IEEEauthorblockA{
    $^1$Department of Computer Science and Software Engineering, Concordia University, Montreal, Canada\\
    $^2$Montreal Neurological Institute, McGill University, Montreal, Canada
  }
}
\maketitle

\begin{abstract}
    Open-access neuroimaging datasets have reached petabyte scale, and continue to grow.
The ability to leverage the entirety of these datasets is limited to a restricted
number of labs with both the capacity and infrastructure to
process the data. Whereas Big Data engines have significantly
reduced application performance penalties with respect to data
movement, their applied strategies (e.g. data locality, in-memory computing and lazy evaluation)
are not necessarily practical
within neuroimaging workflows where intermediary results may
need to be materialized to shared storage for post-processing
analysis. In this paper we evaluate the performance advantage
brought by Intel\textsuperscript{\textregistered}~Optane\texttrademark~DC persistent memory for the
processing of large neuroimaging datasets using the two available
configurations modes: Memory mode and App Direct mode. We
employ a synthetic algorithm on the 76 GiB and 603 GiB BigBrain, as well as apply a
standard neuroimaging application on the Consortium for Reliability
and Reproducibility (CoRR) dataset using 25 and 96 parallel
processes in both cases. Our results show that the performance of
applications leveraging persistent memory is superior to that of other
storage devices, with the exception of DRAM. This is the case
in both Memory and App Direct mode and irrespective of the
amount of data and parallelism. Furthermore, persistent memory in App
Direct mode is believed to benefit from the use of DRAM as
a cache for writing when output data is significantly
smaller than available memory. We believe the use of persistent memory
will be beneficial to both neuroimaging applications running on
HPC or visualization of large, high-resolution images.
\end{abstract}

\section{Introduction}
Neuroimaging open-data initiatives have led to extensively large repositories of
publicly available data. Such initiatives include the \bigbrain~\cite{BigBrain}, 
a one-of-a-kind $\SI{603}{\gibi\byte}$
histological image of a 65 year-old healthy human brain
at 20$\mu$m resolution; the UK Biobank~\cite{ukbiobank}, a repository expected to
contain approximately $\SI{0.2}{\peta\byte}$ of data (including various magnetic
resonance (MR) imaging modalities) from 500,000 individuals living in the UK;
the Human Connectome Project~\cite{HCP}, a repository containing MR scans from
1,200 healthy adults, expected to exceed $\SI{1}{\peta\byte}$ in size; 
and the Consortium for Reliability and Reproducibility (CoRR)~\cite{corr}, an
initiative which aggregates MR data from various centres around the world, 
of which 32 are currently available and make up about $\SI{937}{\gibi\byte}$ of
data in total.

Due to storage limitations, only subsets of such neuroimaging repositories 
can be processed in a typical research laboratory. Moreover, as these datasets 
are extremely large and are only increasing in size, they are typically stored 
in higher-capacity, slower storage devices, such as hard disk drives, or external
parallel file systems, such as Lustre~\cite{lustre}. In such conditions, large-scale 
studies in neuroimaging remain limited to labs with access to adequate 
infrastructures.

Furthermore, intermediary data is often required for post-processing analysis. This limits
any performance benefits that can arise from volatile in-memory computing as intermediary
results need to be materialized onto persistent storage. As neuroimaging datasets continue
to increase in size, the movement of data will significantly increase the processing time.

To mitigate the effects of data writes, the Linux kernel has implemented
strategies, such as the writeback cache. The writeback cache
allows processes to use the memory as a cache for writing. This strategy contrasts writethrough, which 
writes data directly to the device. The writeback cache size
is configurable, but nevertheless limited. When writes approach the cache's capacity,
processes performing writes start to be throttled. Once the cache capacity is reached,
processes can no longer use the cache until all cached written data is flushed
to the appropriate storage device.

Whereas Operating Systems and popular Big Data engines, such as MapReduce~\cite{mapreduce} and Apache Spark~\cite{spark}, have
incorporated software solutions to limit data transfers (e.g. writeback cache, in-memory computing,
data locality, and lazy evaluation), hardware has also adapted to the growing datasets.
One such improvement is the concept of placing persistent storage directly on the Dual
In-line Memory Module (DIMM), thereby reducing the latency of accessing data on 
storage devices. While the latency of these devices is improved, the bandwidth
remains the same. However, as noted by \cite{nvdimms}, a severe performance degradation
can be experienced by having memory traffic and I/O placed on the same bus.

Intel Optane DC Persistent Memory Module~\cite{optanebrief} (DCPMM) is a high-performance
storage technology that resides on the DIMMs to reduce latency to the device.
The first generation of Intel Optane DCPMM offers capabities of 128~GiB,
256~GiB, and 512~GiB, enabling it to be more cost effective than DRAM storage.
There are two configuration modes, Memory mode and App Direct mode,
that enable the storage to either be accessed as an extension of volatile main memory
or as a non-volatile memory storage device~\cite{memory-modes}.

In this paper, we aim to:
\begin{itemize}
        \item Quantify the added value of Intel Optane DC persistent memory on 
            processing large neuroimaging data using representative pipelines; and
        \item Determine when a given Intel Optane DC persistent memory configuration (Memory 
            and App Direct mode) is preferable.
\end{itemize}

\section{Materials and Methods}
The application pipelines, benchmarks, performance data, and analysis scripts used 
to implement the methods described hereafter are all available at 
\url{https://github.com/big-data-lab-team/paper-memory-storage} for 
further inspection and reproducibility.

\subsection{Infrastructure}

The server used consisted of 12$\times$\SI{64}{\gibi\byte} DRAM devices,
resulting in a total of \SI{768}{\gibi\byte} of DRAM, and 12$\times$\SI{256}{\gibi\byte} Optane DCPMM
modules, resulting in total of \SI{3}{\tebi\byte} of Intel Optane DC persistent memory.
Other storage devices included a \SI{240}{\gibi\byte} Micron SATA SSD, of which
only \SI{149}{\gibi\byte} where available for the experiments, and a
\SI{720}{\tebi\byte} shared Dell EMC Isilon network-attached storage platform. Isilon is an NFS 4 mounted
cluster (10Gbps network) consisting of 5 nodes of 36 hard disk drives (HDD). The local disk
was the only storage device set up as a writeback device. Both Isilon and Optane DCPMM in
App Direct mode were configured as write-through devices, meaning they did not leverage
DRAM as a cache for writes. For Optane DCPMM, this was achieved by configuring the XFS filesystem with
direct access (DAX). DAX enables persistent memory to be byte addressable, bypassing the page cache. All storage devices were benchmarked using the script available at \url{https://github.com/big-data-lab-team/paper-memory-storage/blob/master/scripts/bench_disks.sh}.
The result of the benchmarks can be seen in Table \ref{table:bandwidths}.

For processing,
2$\times$Intel(R) Xeon(R) Platinum 8260M CPU @ 2.40GHz where installed, enabling use of up to 96 threads.
The server was running Red Hat Enterprise Linux 7.6 (Maipo) kernel version 3.10.0-957.

\begin{table}
\begin{center}
 \begin{tabular}{ |c|c|c| } 
     \cline{2-3}
     \multicolumn{1}{c|}{} & \multicolumn{2}{c|}{Bandwidth (MB/s)} \\\hline
  Device & Read & Write \\
 \hline
 DRAM & 5304.3 & 3338.2 \\  
 Optane & 3379.2 & 2396.2 \\   
 Local Disk & 518.6 & 240.4 \\
 Isilon & 117.0 & 111.8 \\
 \hline
\end{tabular}\caption{Measured read and write bandwidths}\label{table:bandwidths}
\end{center}
\end{table}

\subsection{Storage configuration}

\subsubsection{Memory mode}

Memory mode leverages Optane DCPMM to extend the system's available
memory. In this mode, Optane DCPMM uses DRAM as a cache and is accessible as
volatile addressable main memory. By extending main memory, Memory mode enables the
fast access of large volumes of data. For instance, Memory mode on our server enabled use of
3~TiB of main memory, whereas it would have only had 768~GiB of main memory if Optane DCPMM could not be
leveraged. Although DRAM is used as cache by Optane DCPMM, it is not visible to the operating system. 

While enabling access to a larger amount of memory than typically accessible otherwise,
it is anticipated that Memory mode will be slower than App Direct mode for all memory accesses that
are not cached in DRAM, due to reduced storage bandwidth. We evaluated all available devices (Optane DCPMM through tmpfs, local SSD and Isilon)
in Memory mode.

\subsubsection{App Direct mode}

App Direct mode enables Optane DCPMM to be accessed as a high-performance storage device.
Unlike Memory mode, the OS is able to differentiate between DRAM and Optane DCPMM,
treating them as two distinct memory tiers. Optane DCPMM
does not use the DRAM as cache in App Direct mode when configured with DAX, as is our case.

Similarly to Memory mode, our experiments evaluated all available filesystems in App Direct
Mode. In other words, we evaluated DRAM (tmpfs mount), Optane DCPMM (\SI{1.5}{\tebi\byte} persistent
memory mount), local SSD and Isilon.

\subsection{Performance model}

We characterize the performance of our data-intensive experiments using the following model:

\begin{equation}
    M \geq \frac{D}{R} + \frac{D}{W} \label{eq:makespan}
\end{equation}

where,
\begin{itemize}
        \item $M$ is the application makespan
        \item $D$ is the total amount of data processed by the application
        \item $R$ is the device read bandwidth
        \item $W$ is the device write bandwidth
\end{itemize}

For applications which have negligible CPU time, as is the case with our experiments, 
it is expected that the I/O duration can estimate the total makespan. However, there
are certain instances where the makespan may be below the I/O duration. This is expected
to occur with scalable devices, as should be the
case for DRAM, Optane DCPMM and Isilon, which can support parallel I/O. Should the
scalable storage devices predict the makespan accurately, it is believed that other overheads
would be at play, making the device unfavourable for parallel processing.

\subsection{Applications}
\subsubsection{ \bigbrain Incrementation}

\begin{algorithm}\caption{Incrementation}\label{alg:incrementation}
    \begin{algorithmic}[1]
    \Input
        \Desc{$x$}{a sleep delay in seconds}
        \Desc{$n$}{a number of iterations}
        \Desc{$C$}{a set of image chunks}
        \Desc{$fs$}{filesystem to write to (tmpfs, Optane DCPMM, local disk, Isilon)}
    \EndInput
    \ForEach{$chunk \in C$}
        \State read $chunk$ from $fs$
        \State $chunk\gets chunk+1$
        \State save $chunk$ to $fs$
    \EndFor
    \end{algorithmic}
\end{algorithm}  

Due to the size and uniqueness of the \bigbrain, standardized processing pipelines
have yet to be developed. In order to quantify the effects of storage devices
on such a dataset, we have implemented a basic synthetic application that takes 
the image, split into blocks, and increments all the voxels within each block, in parallel (Algorithm~\ref{alg:incrementation}).
This application
enables us to read and write to different storage devices and ensure that written
data could not have been previously cached
in-memory, by ensuring that read and written data are not the same. This application
was parallelized in two different ways: using both GNU Parallel~\cite{gnuparallel} and Apache Spark 2.4.3 (PySpark)~\cite{spark}. 
All code was implemented in Python 3.6.

There are some notable differences between the GNU Parallel and Apache Spark implementations.
For instance, in the GNU Parallel implementation, all operations (i.e. reading, incrementing and writing)
are done within a single task. This task is applied to all the \bigbrain blocks using GNU Parallel,
processing a subset of them at a time,
depending on the level of parallelism provided. Reading in the GNU Parallel implementation 
is achieved using the popular neuroimaging I/O library NiBabel v2.5.0~\cite{nibabel}. When provided with a filename, as is the case
for the GNU Parallel implementation, Nibabel will simply load the header in memory, and memory map the data
using \href{https://numpy.org/}{NumPy}. This step will be referred to as ``load header''. It is only when the data is actually required
(i.e. during incrementation), that the data will be loaded into memory.

The Apache Spark implementation differs from that of GNU Parallel. Whereas GNU Parallel will
simply forks processes for each task and defers scheduling to the kernel, Spark is responsible for 
task scheduling decisions. Furthermore, in our Spark implementation, we opted to read the data using Spark's built-in
\texttt{BinaryFiles}, loading whole files, in binary format, into Spark partitions. As the
data would have been pre-loaded by Spark, ``loading header'' only measures the time to
convert the binary images into NiBabel objects, and ``incrementation'' consists solely
of the time it took to increment the data. Therefore, only the write time would be
comparable between the two implementations, as it is achieved using the same method.
Furthermore, the Spark implementation also differs in that read, increment and write
were separated into three map tasks, which can enable shuffling between the tasks should 
it be determined as necessary by the Spark scheduler.

Time is measured within the application using Python's \texttt{time} module 
called before and after any read and writes.

We have executed this pipeline on both the $\SI{75}{\gibi\byte}$ 40~$\mu$m 
\bigbrain split into 125$\times$$\SI{614}{\mebi\byte}$ blocks and the $\SI{603}{\gibi\byte}$
\bigbrain at 20~$\mu$m split into 1000$\times$$\SI{617}{\mebi\byte}$ blocks.

For the 40~$\mu$m \bigbrain, we have executed the application using GNU Parallel for parallelization
using 25 and 96 processes on the 125 40$\mu$m \bigbrain blocks. Data was read and written to either
DRAM (App Direct only), Optane DCPMM, local disk and Isilon. 

For the 20~$\mu$m \bigbrain, we used the same application, however in this case, it was executed
using both Apache Spark and GNU Parallel. The configuration was also generally the same, having 
experiments using both 25 and 96 processes. Repeating the experiments with the 20~$\mu$m BigBrain would enable
us to determine the effects when Optane DCPMM would have to be partially relied upon due to insufficient DRAM
space. Moreover, with such large datasets, it is likely that Big Data frameworks would be used rather than
more traditional parallelism frameworks. Using Spark, we can evaluate how Big Data Frameworks perform on
different storage devices.

Since storage was limited on DRAM and local disk and was not large enough to process the entire
20~$\mu$m \bigbrain, these devices were omitted in the processing of this dataset. 

\subsubsection{BIDS App Example}

The BIDS App Example is a template example for creating a Brain Imaging Data Structure (BIDS)
compliant application. It runs a standard neuroimaging brain extraction application 
on all the anatomical images of datasets containing numerous subjects. This step is 
referred to as Participant analysis within the application. An optional
step of the BIDS App example, referred to as Group analysis, computes the average brain
mask size of the entire dataset.

We used the entire CoRR dataset, available on \href{https://www.datalad.org/}{DataLad} and
applied Participant analysis to it. The BIDS App Example was executed using a Singularity container stored on
Isilon. As in the BigBrain Incrementation, the experiment was parallelized 
using GNU Parallel with 25 and 96 processes. The conditions were executed in both App Direct and Memory mode,
where Isilon and the local SSD were only evaluated in Memory mode. Each experiment was repeated 3x.

Time in this application was obtained using Linux's \texttt{time()} application.
\texttt{Real + Sys} was used to measure CPU time,
whereas \texttt{User - (Real + Sys)} was used to measure
I/O time.
\section{Results}

\subsection{40~$\mu$m \bigbrain incrementation}
\subsubsection{25 Processes}

\begin{figure}
    \begin{subfigure}{\columnwidth}
        \centering
        \includegraphics[width=\columnwidth]{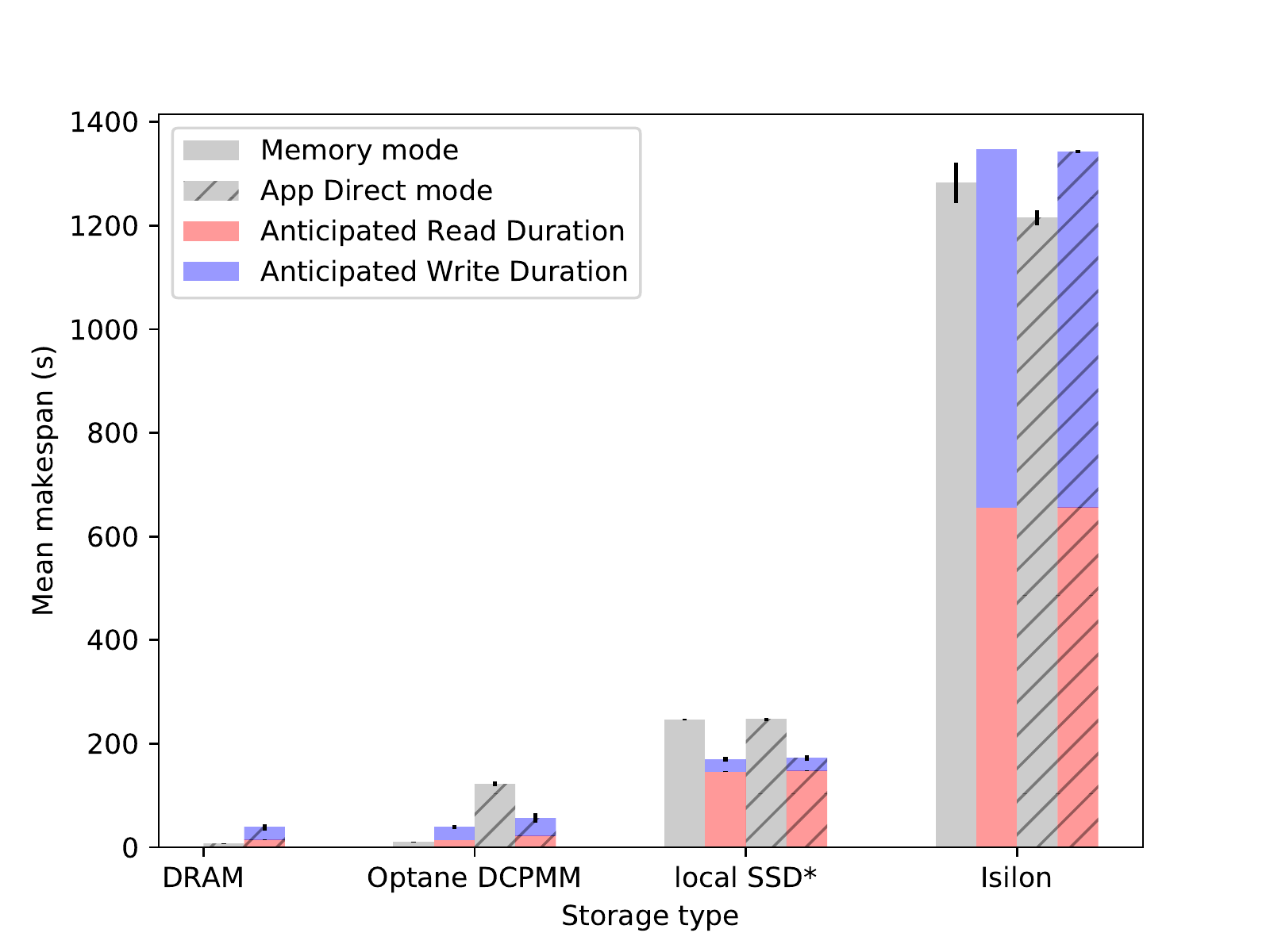}
        \caption{Makespan}\label{fig:makespan-25cpus}
    \end{subfigure}
    \begin{subfigure}{\columnwidth}
        \centering
        \includegraphics[width=\columnwidth]{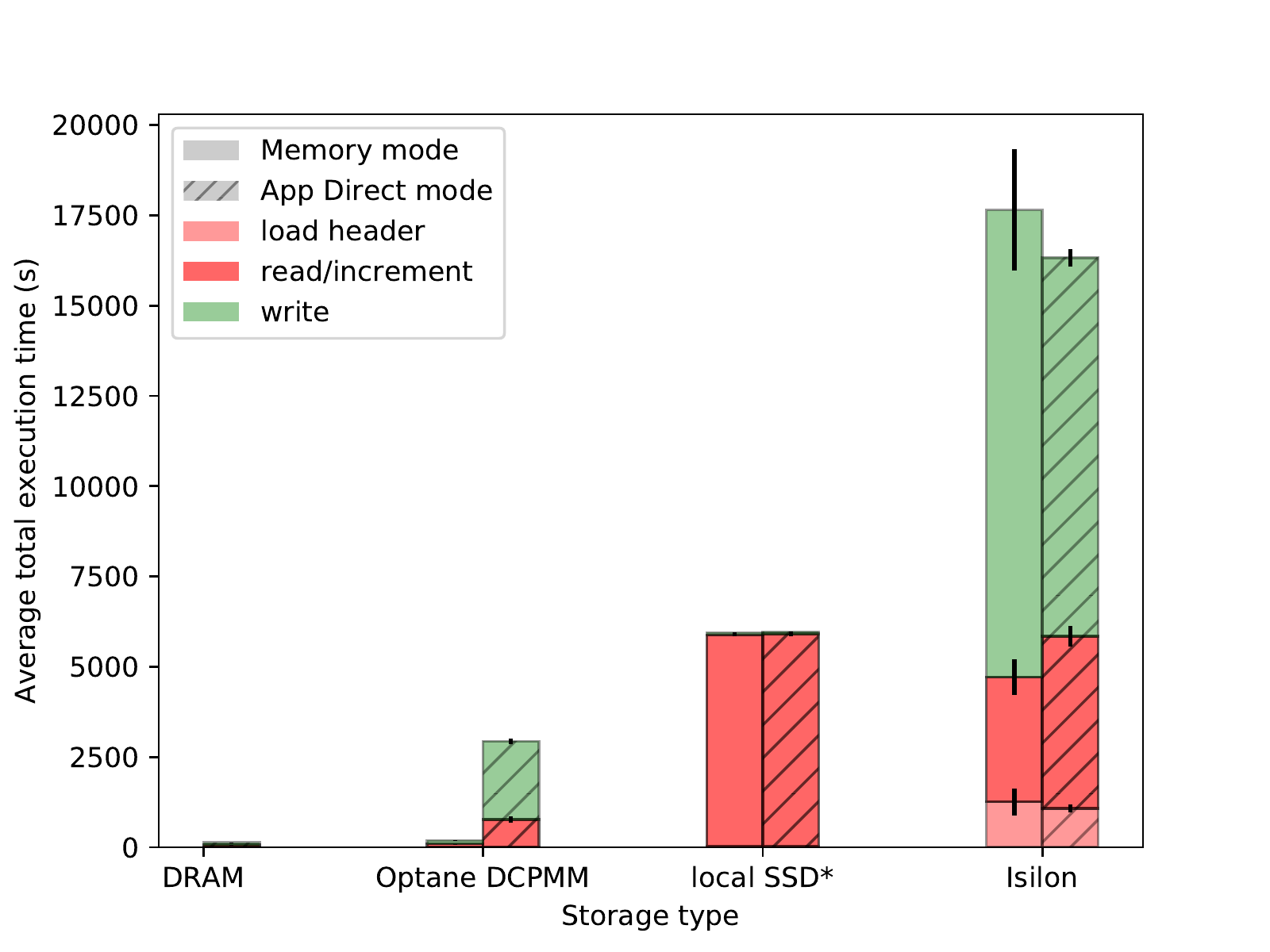}
        \caption{Total read/increment/write breakdown}\label{fig:stacked-25cpus}
    \end{subfigure}
    \captionsetup{belowskip=-10pt}
    \caption{GNU Parallel incrementation application processing the 40 $\mu$m BigBrain using 25 processes (three repetitions). Anticipated read and write duration was 
    measured by applying the perceived bandwidths Table~\ref{table:bandwidths}
    to Eq.~\ref{eq:makespan}. *Local SSD did not complete the
    writing of the last few blocks (approximately 5-10) due to storage limitations.}
\end{figure}

\begin{figure*}
    \begin{subfigure}{\columnwidth}
        \centering
    \includegraphics[width=\columnwidth]{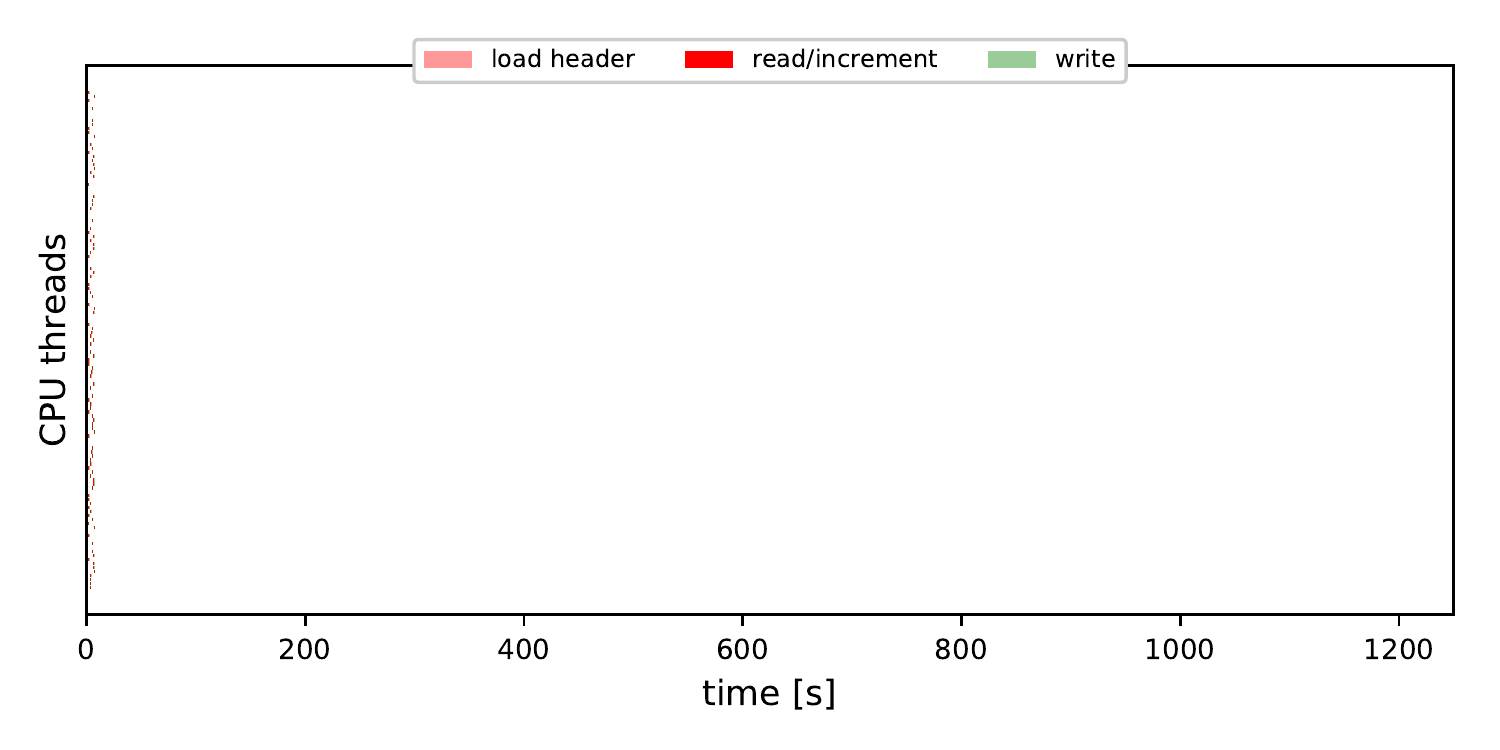}
    \caption{DRAM}
\end{subfigure}
\begin{subfigure}{\columnwidth}
        \centering
    \includegraphics[width=\columnwidth]{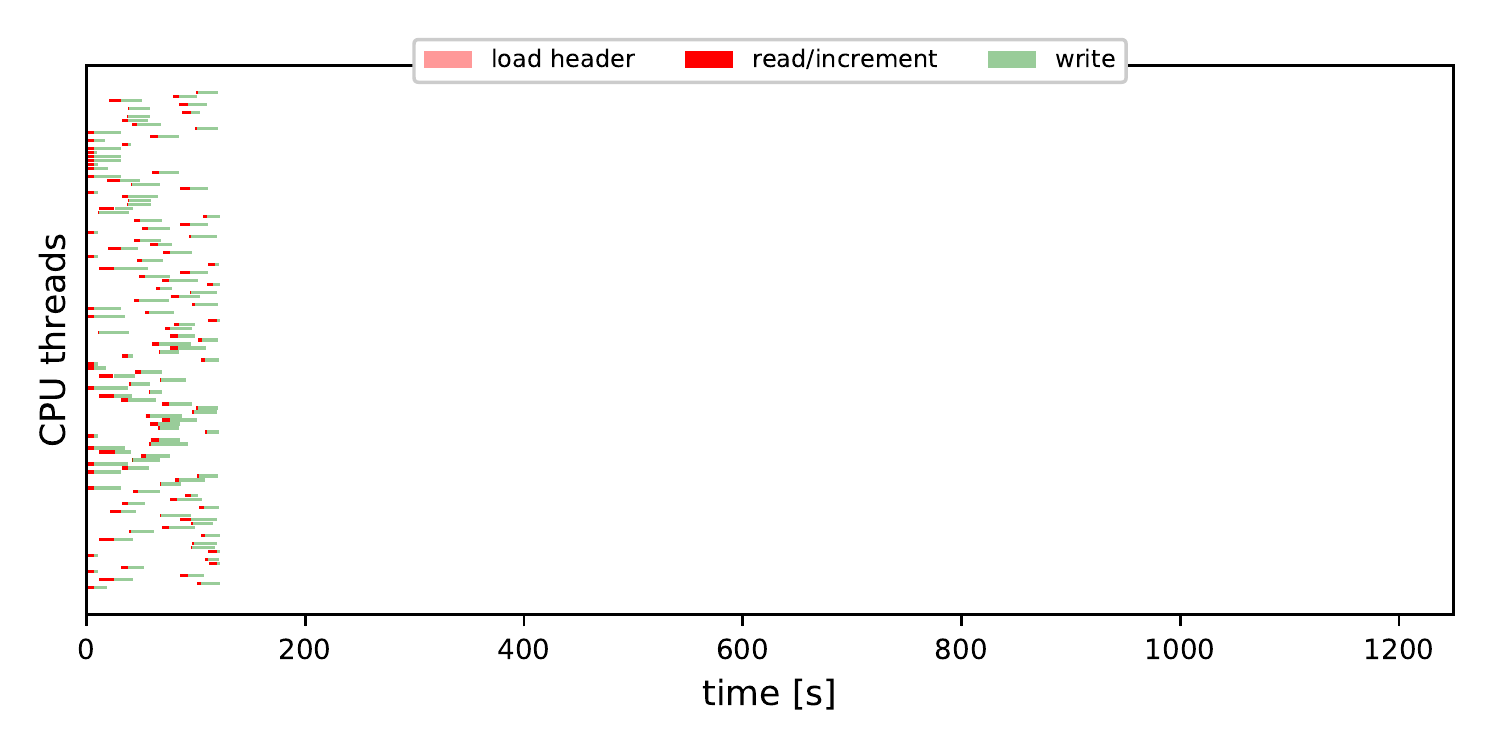}
    \caption{Optane DCPMM}
\end{subfigure}
\begin{subfigure}{\columnwidth}
        \centering
    \includegraphics[width=\columnwidth]{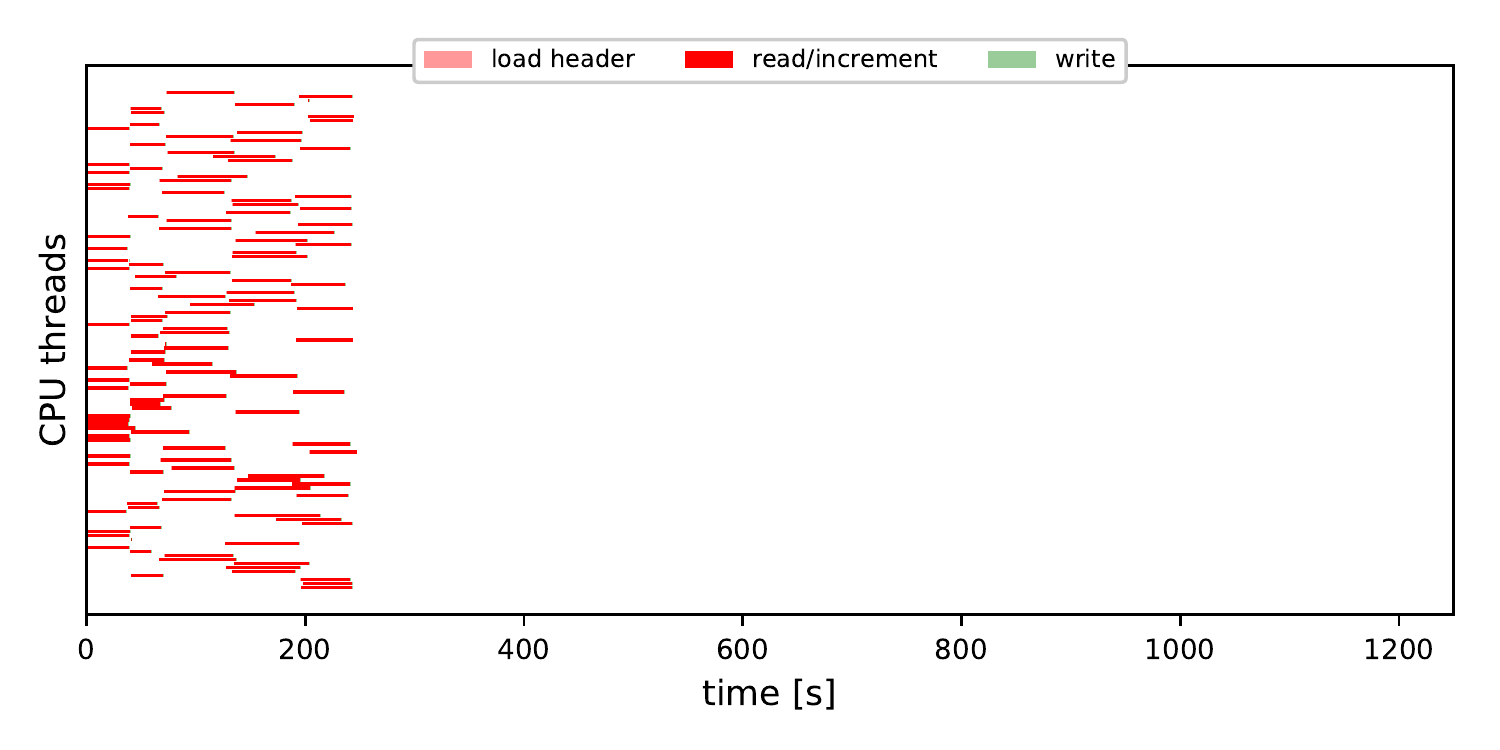}
    \caption{local SSD*}
\end{subfigure}
\begin{subfigure}{\columnwidth}
        \centering
    \includegraphics[width=\columnwidth]{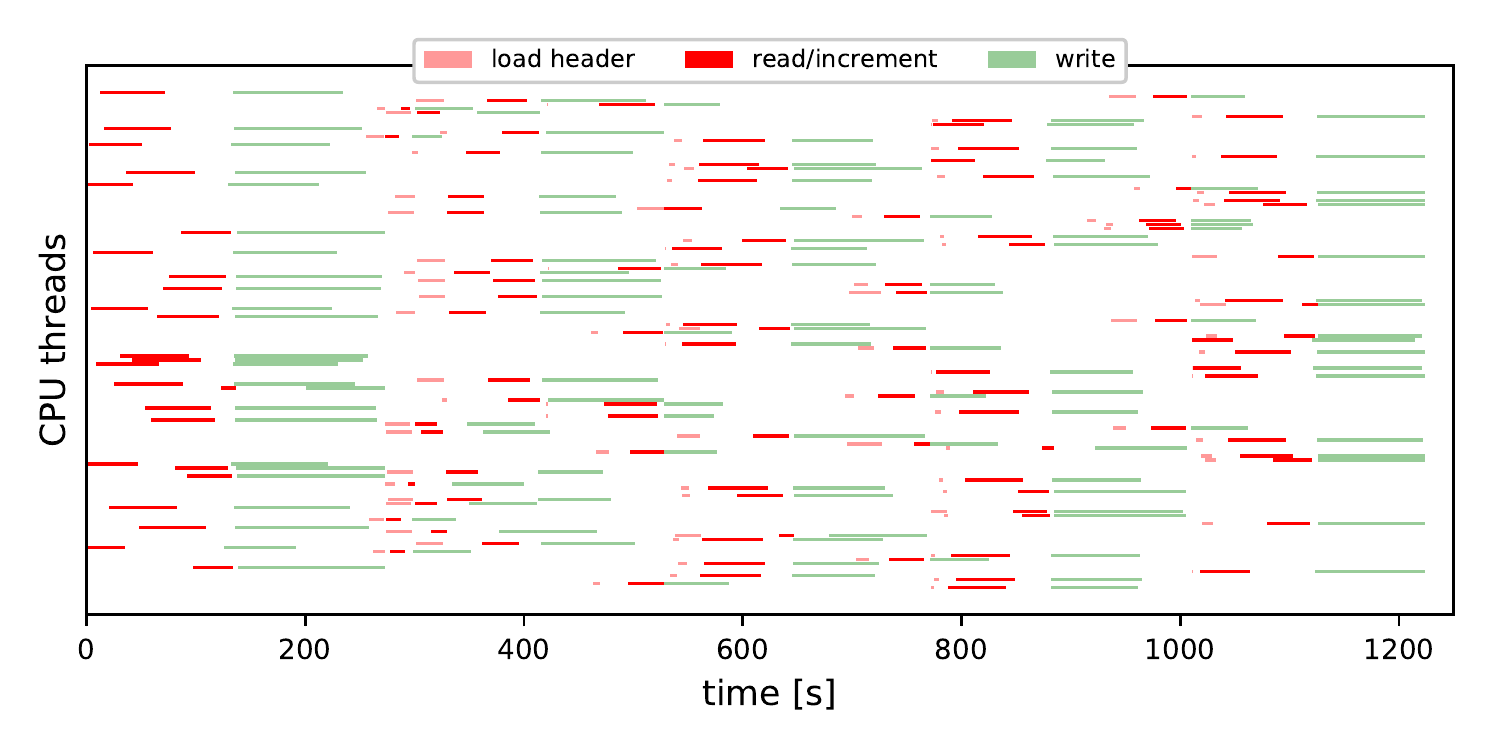}
    \caption{Isilon}\label{fig:gantt25isilon}
\end{subfigure}
    \captionsetup{belowskip=-10pt}
    \caption{Gantt charts for each storage device (App Direct mode) processing 125 blocks of the 40$\mu$m BigBrain using 25 processes. *Some local SSD writes (approximately 5-10 writes) did not complete due to storage limitations.}\label{fig:gantt25}
\end{figure*}

As can be seen in Figure~\ref{fig:makespan-25cpus}, DRAM (App Direct mode) is the 
most efficient, with a makespan of 7.6s (all makespan values are available \href{https://github.com/big-data-lab-team/paper-memory-storage/blob/master/experiments/bigbrain-incrementation/results/makespan.csv}{here}).
Optane DCPMM in Memory mode is a close second
with a makespan of 10.2s. As caching was disabled in App Direct mode, 
Optane DCPMM in this mode is approximately 
12x slower than Optane DCPMM in Memory mode. Isilon in both Memory mode and App Direct 
mode is more than 120x slower than Optane DCPMM in Memory mode. Other than for Optane DCPMM, 
there is no significant difference between Memory mode and App Direct executions.

When comparing the makespan to the estimated read and write duration, local SSD
in both modes and Optane DCPMM in App Direct mode have longer makespans than anticipated.
Optane DCPMM was approximately 2.2x slower than expected, whereas local SSD was approximately
1.4x slower than expected, in both modes. Conversely,
DRAM was almost 5x faster in reality compared to what was estimated. Optane DCPMM in Memory mode
had a larger difference, with it being 3.8x faster than expected. Isilon in App Direct
mode had a smaller difference than Memory mode, with it being 1.1x faster than expected.

The total task duration breakdowns for each device (Figure~\ref{fig:stacked-25cpus})
shows that, as expected, I/O times vary greatly between devices, with DRAM having the
best total read and write speeds. Optane DCPMM in Memory mode is very close to DRAM in speed,
whereas Optane DCPMM in App Direct mode was 7x slower in terms of reading and 23x slower in terms
of writing. Local SSD, on the other hand, exhibits similar write times as DRAM, but is
approximately 96x slower than DRAM with respect to reads. Header loading for local SSD is also 90x longer
than that of DRAM, however, it remains negligible compared to other tasks. Isilon, on the 
other hand, has non negligible header load times, with it being more than 5000x
slower than that of DRAM. Interestingly enough, the read/increment times on Isilon are
faster than that of the local SSD, although Isilon makes up for its speedier reads with 
writes that are around 184x slower than DRAM.

The App Direct mode Gantt charts (Figure~\ref{fig:gantt25}) also reflect what is observed in Figure~\ref{fig:stacked-25cpus}.
DRAM is significantly faster than all other storage, leading it to barely appear within the Gantt chart.
DRAM was measured to have an average of 17.2 parallel tasks throughout the execution. Optane DCPMM takes
significantly longer than DRAM and appears to spend a significant amount of time writing. The average
parallelism measured for Optane DCPMM was 24. Unlike Optane DCPMM, local SSD spends the vast majority of its time
reading, while spending a negligible amount of time writing, due to writeback to DRAM. Like Optane DCPMM, it averages 24 parallel tasks
throughout its execution. Unlike the other three storage, Isilon shows spacing between the loading of the header,
read and increment and write. Isilon was measured to have an average parallelism of 14. Overall, each storage device
execution displayed heterogeneous task durations.

\subsubsection{96 Processes}

\begin{figure}
    \begin{subfigure}{\columnwidth}
        \centering
        \includegraphics[width=\columnwidth]{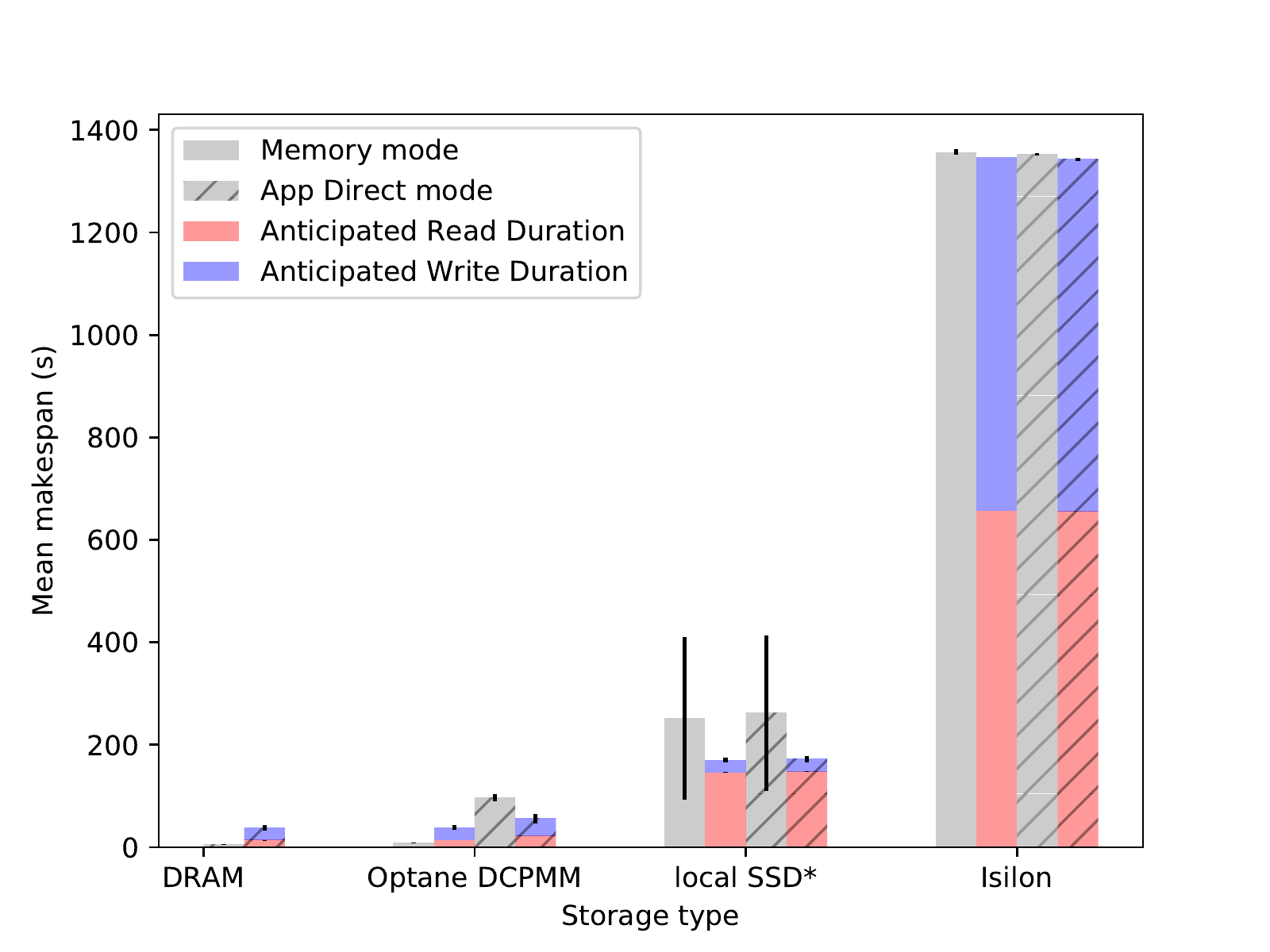}
        \caption{Makespan}\label{fig:makespan-96cpus}
    \end{subfigure}
    \begin{subfigure}{\columnwidth}
        \centering
        \includegraphics[width=\columnwidth]{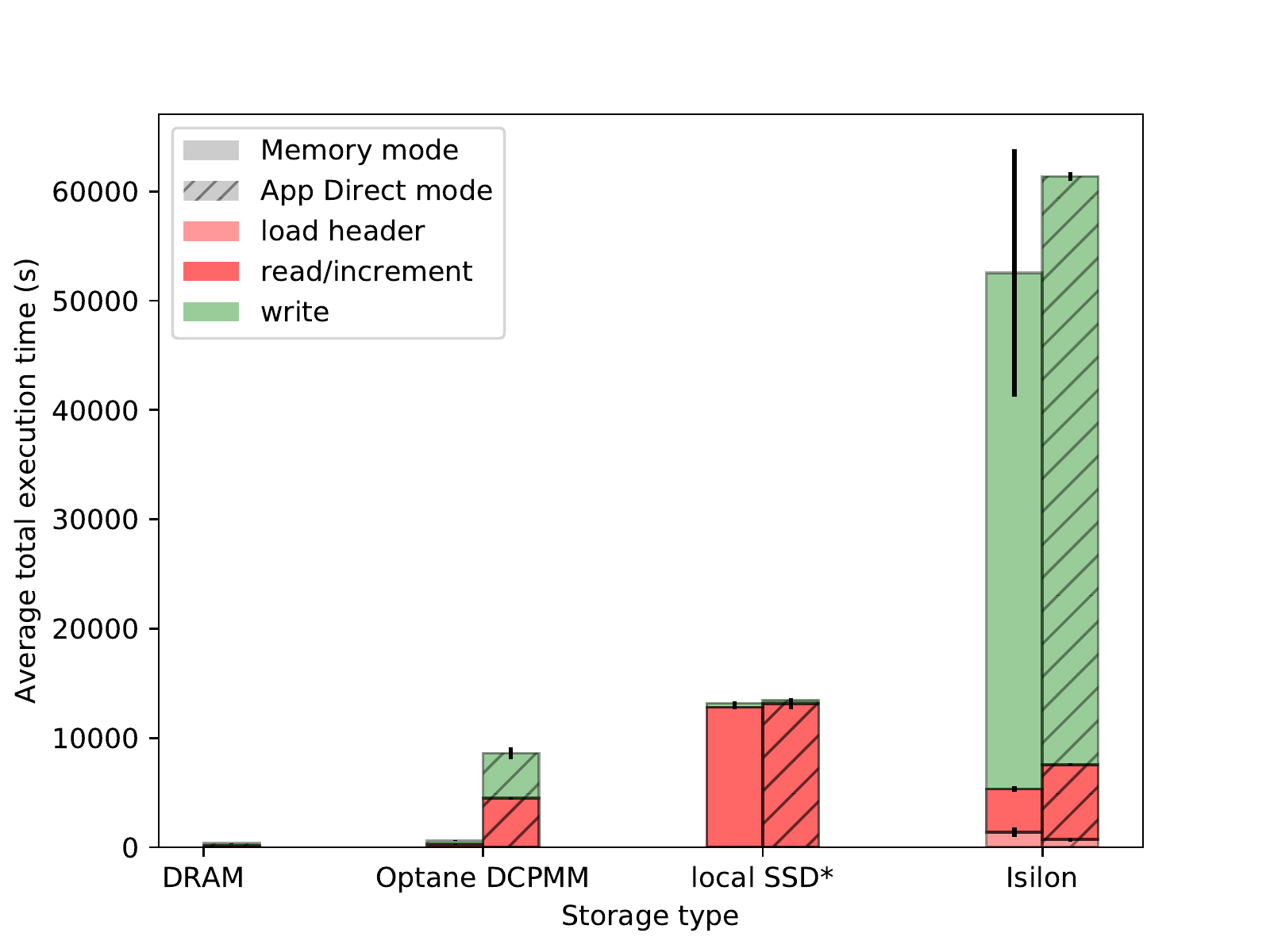}
        \caption{Total read/increment/write breakdown}\label{fig:stacked-96cpus}
    \end{subfigure}
    \captionsetup{belowskip=-10pt}
    \caption{GNU Parallel incrementation application processing the 40 $\mu$m BigBrain using 96 processes (three repetitions). *Local SSD did not complete the
    writing of the last few blocks (approximately 5-10) due to storage limitations.}
\end{figure}

\begin{figure*}
    \begin{subfigure}{\columnwidth}
        \centering
    \includegraphics[width=\columnwidth]{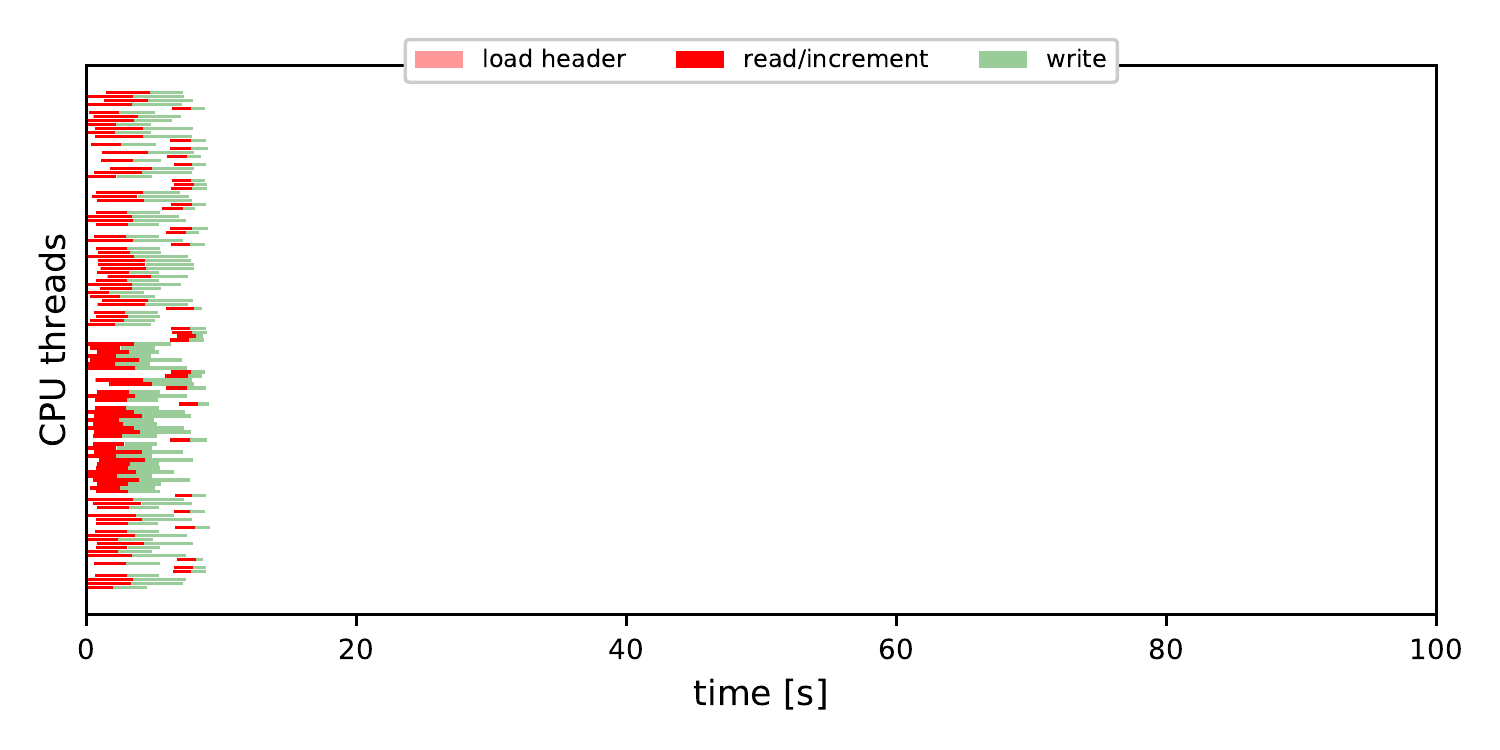}
    \caption{Memory mode}
\end{subfigure}
\begin{subfigure}{\columnwidth}
        \centering
    \includegraphics[width=\columnwidth]{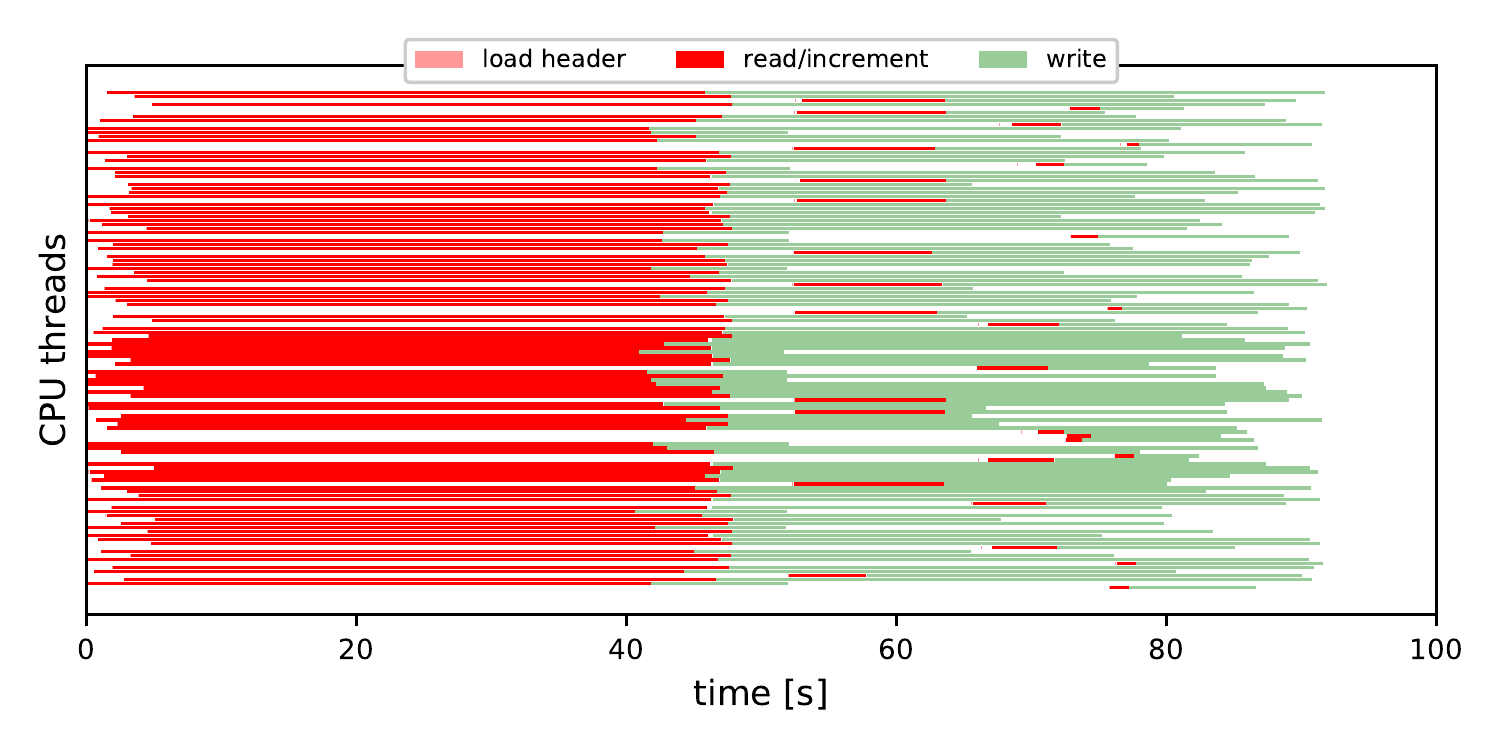}
    \caption{App Direct mode}
\end{subfigure}
    \captionsetup{belowskip=-10pt}
\caption{Gantt charts for Optane DCPMM processing 125 blocks of the 40$\mu$m BigBrain using 96 processes}\label{fig:gantt96}
\end{figure*}

Despite having increased parallelism by a factor of 3.84, we see no visible reduction
in makespan (Figure~\ref{fig:makespan-96cpus}). At 96 processes, DRAM and Optane DCPMM in Memory mode are the only devices that did better than the anticipated I/O, with a makespan duration of
5.7s and 9s, respectively (total read and write estimate for both was measured to be around 39s). While
Optane DCPMM in App Direct mode and local SSD both performed worse than the anticipated I/O estimates, Isilon
performed as expected. Variance was found to be high on the local SSD.

The total task duration breakdowns (Figure~\ref{fig:stacked-96cpus}) show that total task durations
have nearly quadrupled. Whereas most devices decreased in performance in both reading and writing,
Isilon only appeared to display a performance decrease with respect to writes. Furthermore, while both
read and writes decreased for Optane DCPMM in App Direct mode, read duration was nearly 6x slower with more threads,
while write duration was only about 2x slower. For local SSD and Isilon, there does not appear to 
be any significant difference between App Direct and Memory mode.

When analyzing the average parallelism, only Optane DCPMM in App Direct mode
came close to 96 parallel tasks, with an average parallelism of 88 tasks. Optane DCPMM in Memory mode,
DRAM and local SSD all averaged between 61-66 tasks. Isilon performed the worst with 38 tasks in Memory 
mode and 45 tasks in App Direct mode.

A further look at the Gantt charts between Optane DCPMM in Memory mode and App Direct mode (Figure~\ref{fig:gantt96})
shows that both read and write time of tasks are worse in App Direct mode as compared to Memory mode.
While some read tasks in App Direct mode appear to be of similar duration to the average read task time
of Memory mode, the vast majority of read tasks are significantly slower. Furthermore, no write tasks in App
Direct mode is capable of reaching the performance of equivalent tasks in Memory mode.

\subsection{20~$\mu$m \bigbrain incrementation}
\subsubsection{25 Processes}

\begin{figure*}
    \begin{subfigure}{\columnwidth}
        \centering
    \includegraphics[width=\columnwidth]{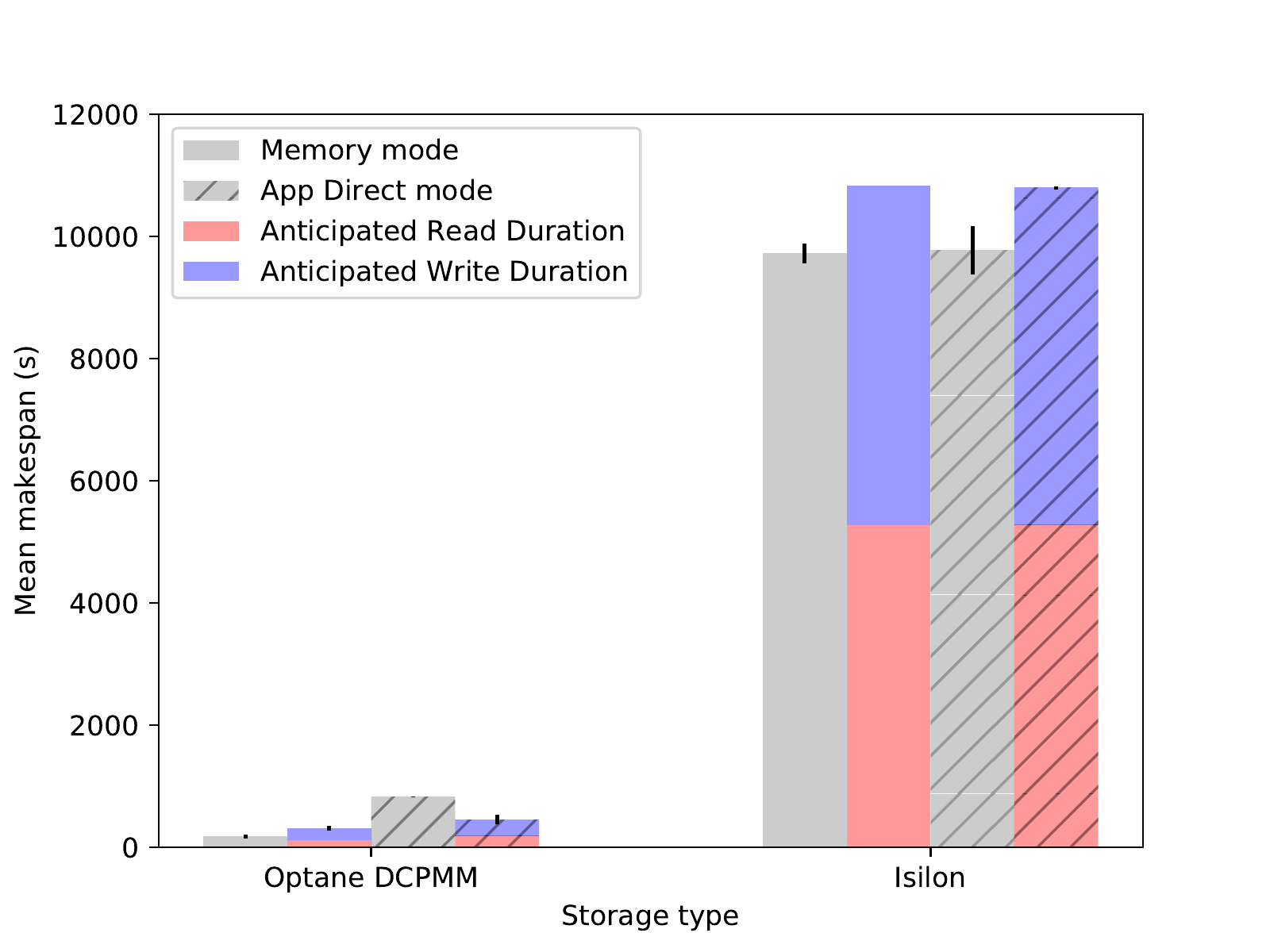}
    \caption{Makespan}\label{fig:20mksp25}
\end{subfigure}
\begin{subfigure}{\columnwidth}
        \centering
    \includegraphics[width=\columnwidth]{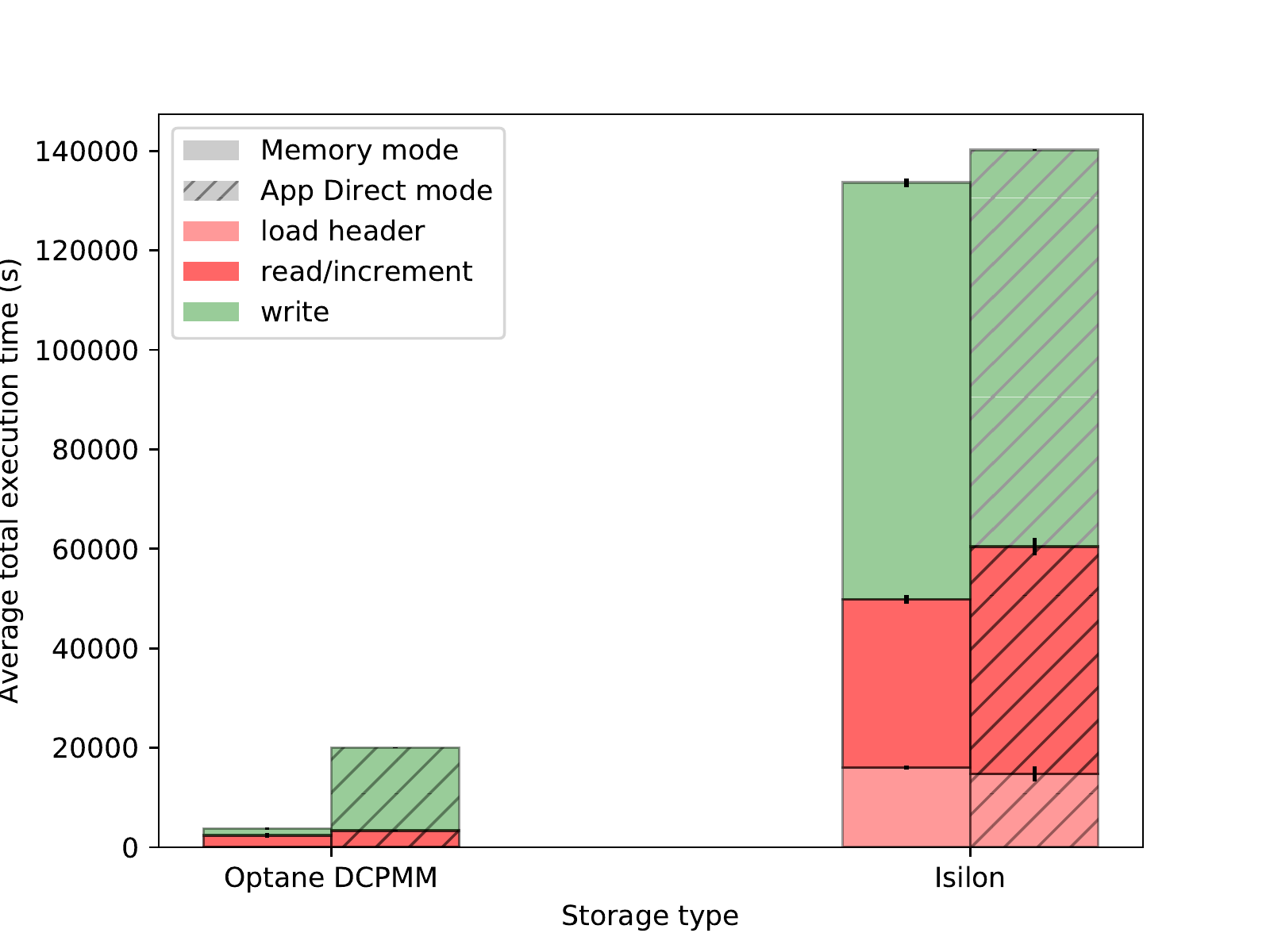}
    \caption{Total read/increment/write breakdown}\label{fig:20total25}
\end{subfigure}
\captionsetup{belowskip=-10pt}
\caption{GNU Parallel incrementation application processing the 20~$\mu$m BigBrain using
25 processes (three repetitions).}\label{fig:2025}
\end{figure*}

The anticipated I/O estimates did not correctly predict the makespan for 25 parallel
processes (Figure~\ref{fig:20mksp25}). Both Optane DCPMM in Memory mode and Isilon appear 
to be under the estimates: Optane DCPMM in Memory mode is approximately twice as fast as the
estimates whereas Isilon is around 1.1x faster. Optane was 1.8x slower than anticipated.
Once again, Memory and App Direct mode appear to only affect performance for Optane.

Total task duration breakdowns (Figure~\ref{fig:20total25}) show that Optane DCPMM in
App Direct mode spends significantly more time writing than in Memory mode.
In fact, it spends almost 12x more time writing. Isilon also appears to vary slightly
between Memory and App Direct mode, spending around 1.4x more time reading and incrementing
in App Direct mode. 

\begin{figure*}
    \begin{subfigure}{\columnwidth}
        \centering
    \includegraphics[width=\columnwidth]{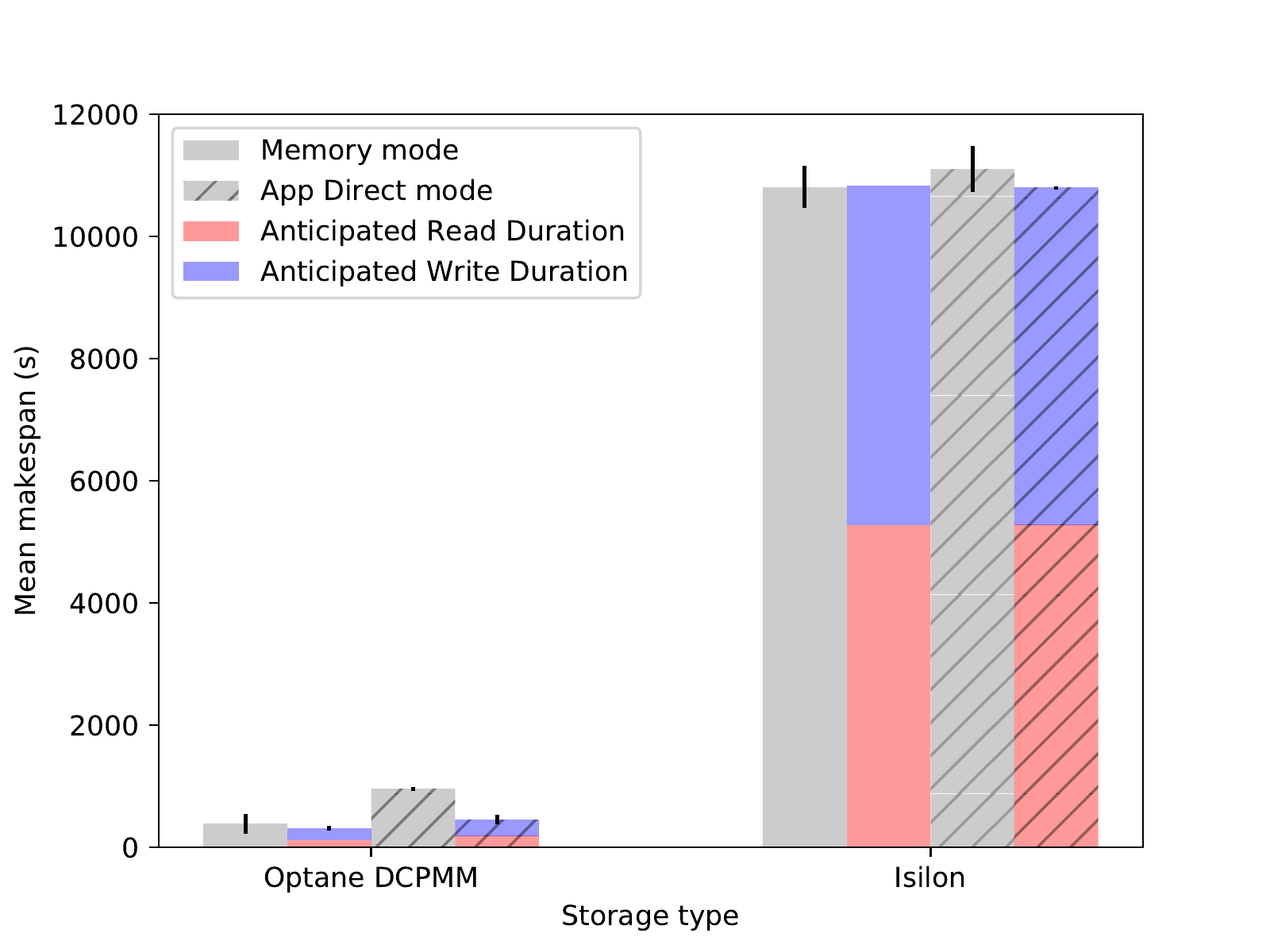}
    \caption{Makespan}\label{fig:20mkspspark25}
\end{subfigure}
\begin{subfigure}{\columnwidth}
        \centering
    \includegraphics[width=\columnwidth]{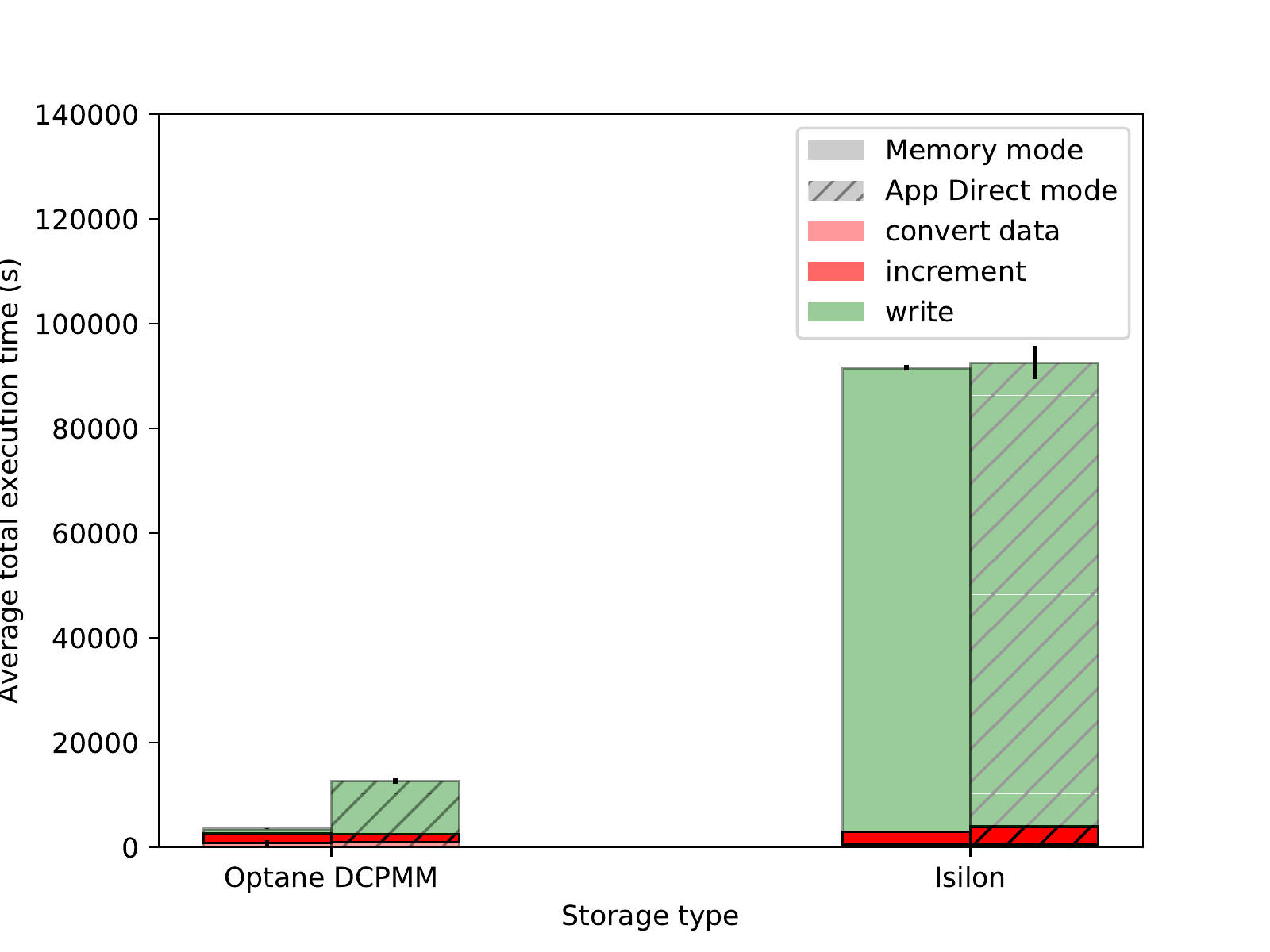}
    \caption{Total convert/increment/write breadown}\label{fig:20totalspark:25}
\end{subfigure}
\caption{Spark incrementation application processing the 20~$\mu$m BigBrain using 25 processes (three repetitions).}\label{fig:20spark25}
\captionsetup{belowskip=-10pt}
\end{figure*}

Unlike the GNU Parallel executions, the Spark executions appear to be in better accordance with
the I/O estimates (Figure~\ref{fig:20mkspspark25}). Only Optane DCPMM in App Direct mode appears
to have significantly exceeded the estimates, by a factor of 2. Total write duration all appear
to be longer in the Spark implementation than with GNU Parallel (Figure~\ref{fig:20totalspark:25}).
Spark's writes are 1.4-1.6x longer with Optane DCPMM and 1.1x longer with Isilon. Due to differences in
how data is loaded in Spark, it is not possible to comment on how read times are affected. However, it
is possible to note that data conversion (binary string to NumPy array) takes almost twice as 
much time on Optane DCPMM than it does on Isilon. Furthermore, the act of incrementation can take
up to twice as much time on Isilon.

\subsubsection{96 Processes}

\begin{figure*}
    \begin{subfigure}{\columnwidth}
        \centering
    \includegraphics[width=\columnwidth]{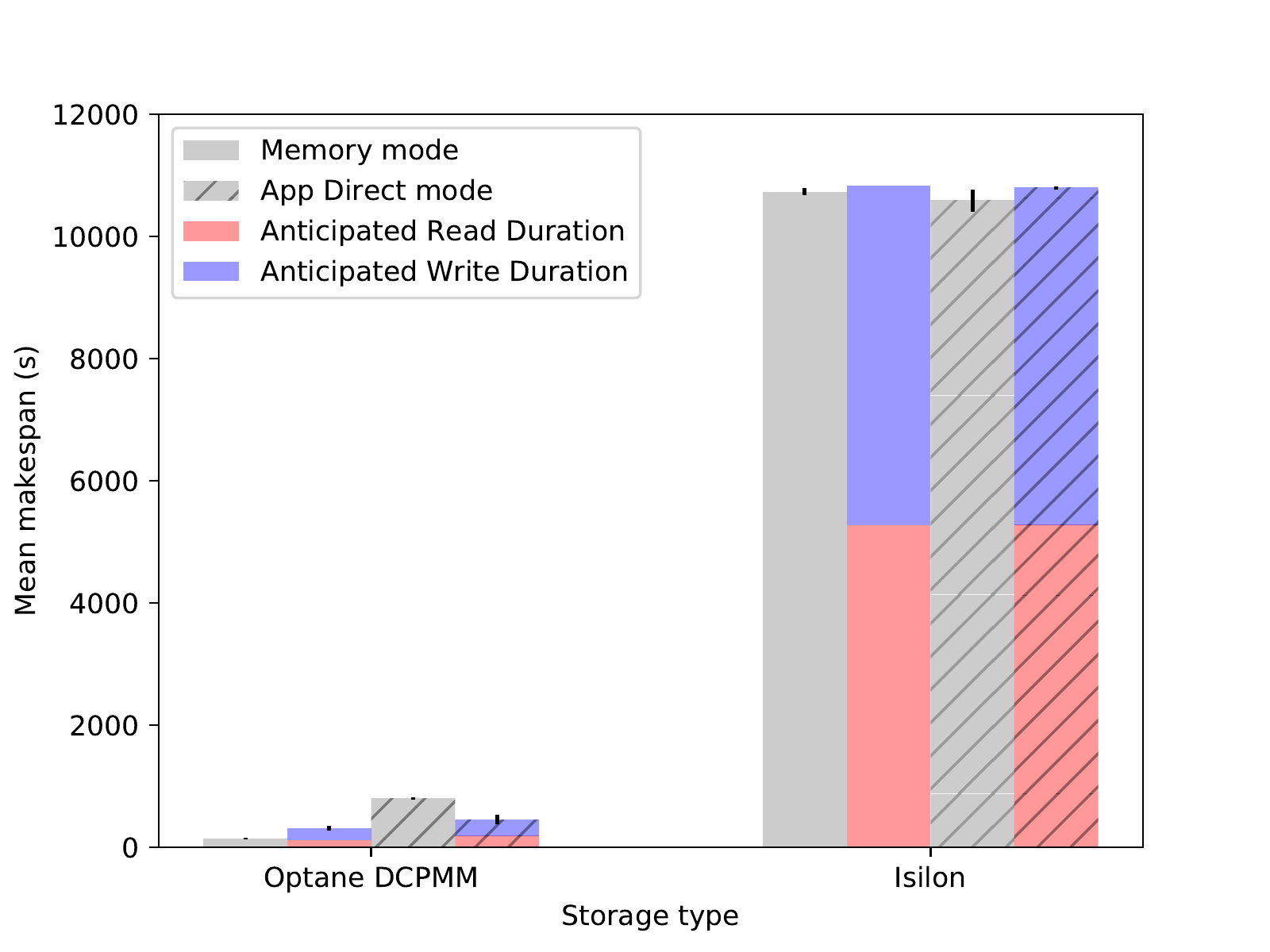}
    \caption{Makespan}\label{fig:20mksp96}
\end{subfigure}
\begin{subfigure}{\columnwidth}
        \centering
    \includegraphics[width=\columnwidth]{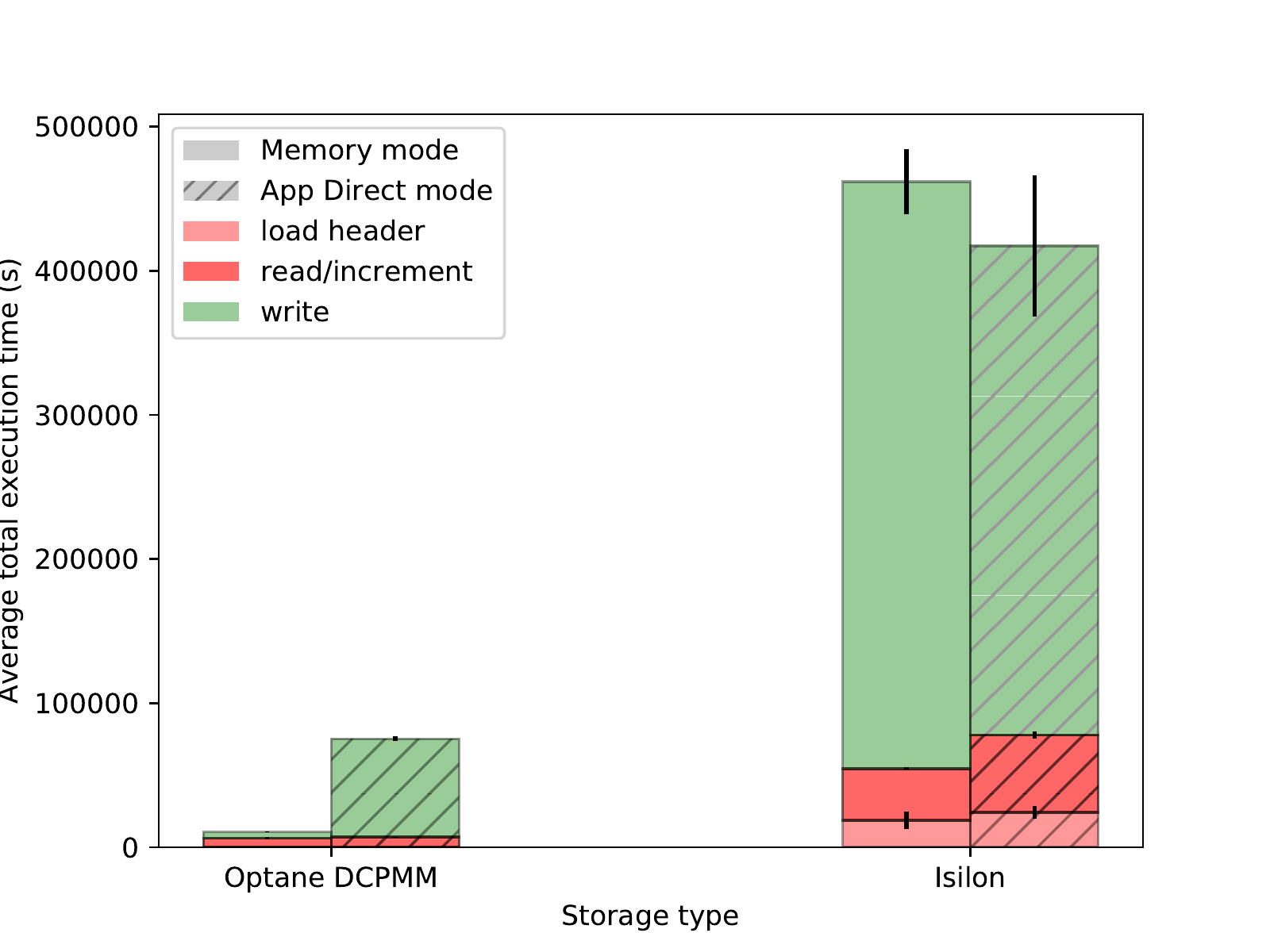}
    \caption{Total read/increment/write breakdown}\label{fig:20total96}
\end{subfigure}
\caption{GNU Parallel incrementation application processing the 20~$\mu$m BigBrain using
96 processes (three repetitions).}\label{fig:2096}
\captionsetup{belowskip=-10pt}
\end{figure*}

A very slight performance improvement (1-1.3x faster) can be witnessed in Optane DCPMM (Figure~\ref{fig:20mksp96}). In 
contrast, Isilon 
experienced a slight performance decrease (1.1x slower). Whereas Isilon makespan appears to match
estimated I/O times, Optane DCPMM once again has a superior Memory mode makespan and an inferior App Direct
mode makespan (Figure~\ref{fig:20mksp96}). As previously observed with 25 processes,
the bulk of the processing time is taken up by writing for Optane DCPMM in App Direct mode and 
Isilon in both modes (Figure~\ref{fig:20total96}). Reading takes more time in the case of Optane DCPMM 
in Memory mode.

\begin{figure*}
    \begin{subfigure}{\columnwidth}
        \centering
    \includegraphics[width=\columnwidth]{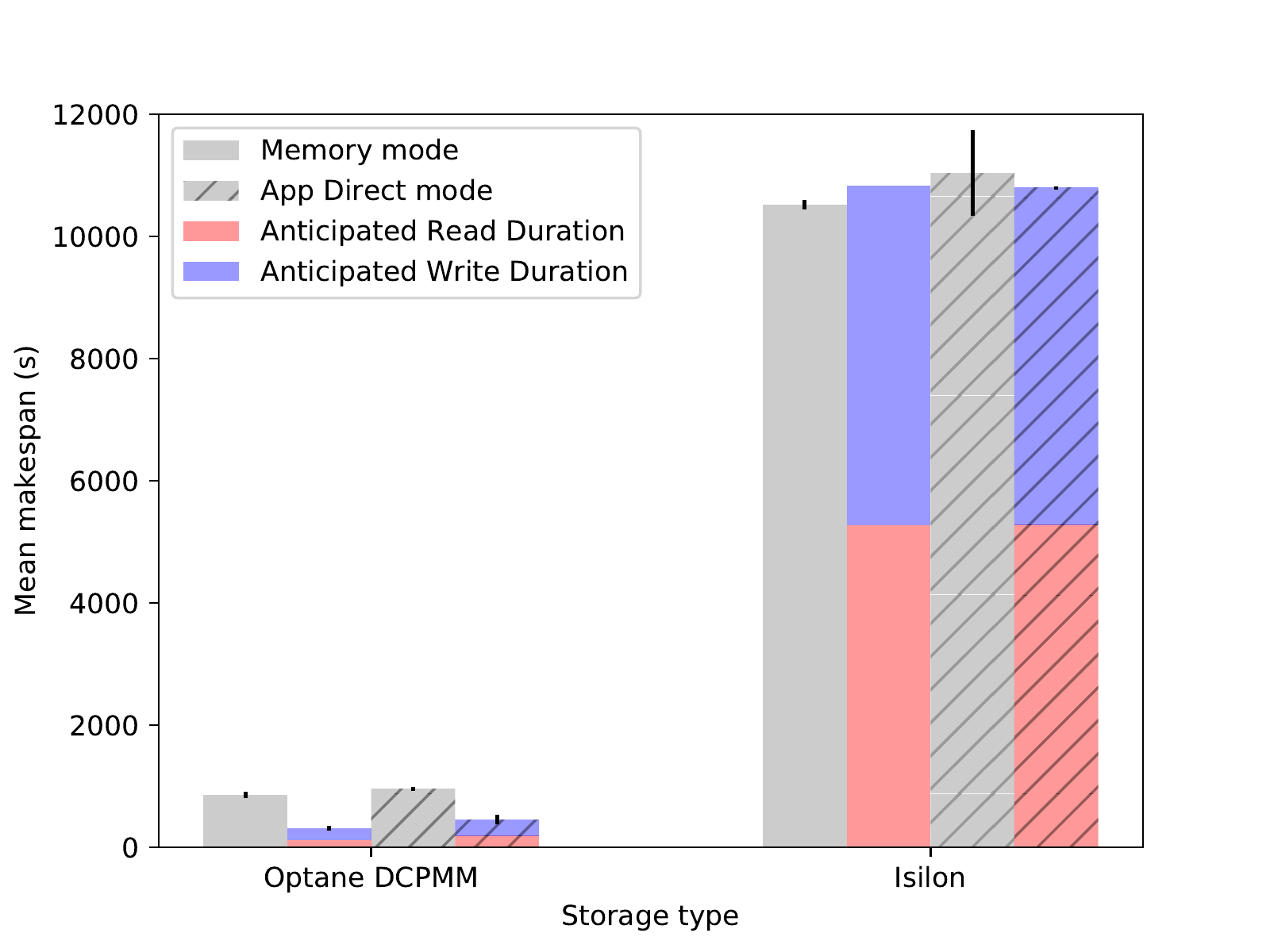}
    \caption{Makespan}\label{fig:20mkspspark96}
\end{subfigure}
\begin{subfigure}{\columnwidth}
        \centering
    \includegraphics[width=\columnwidth]{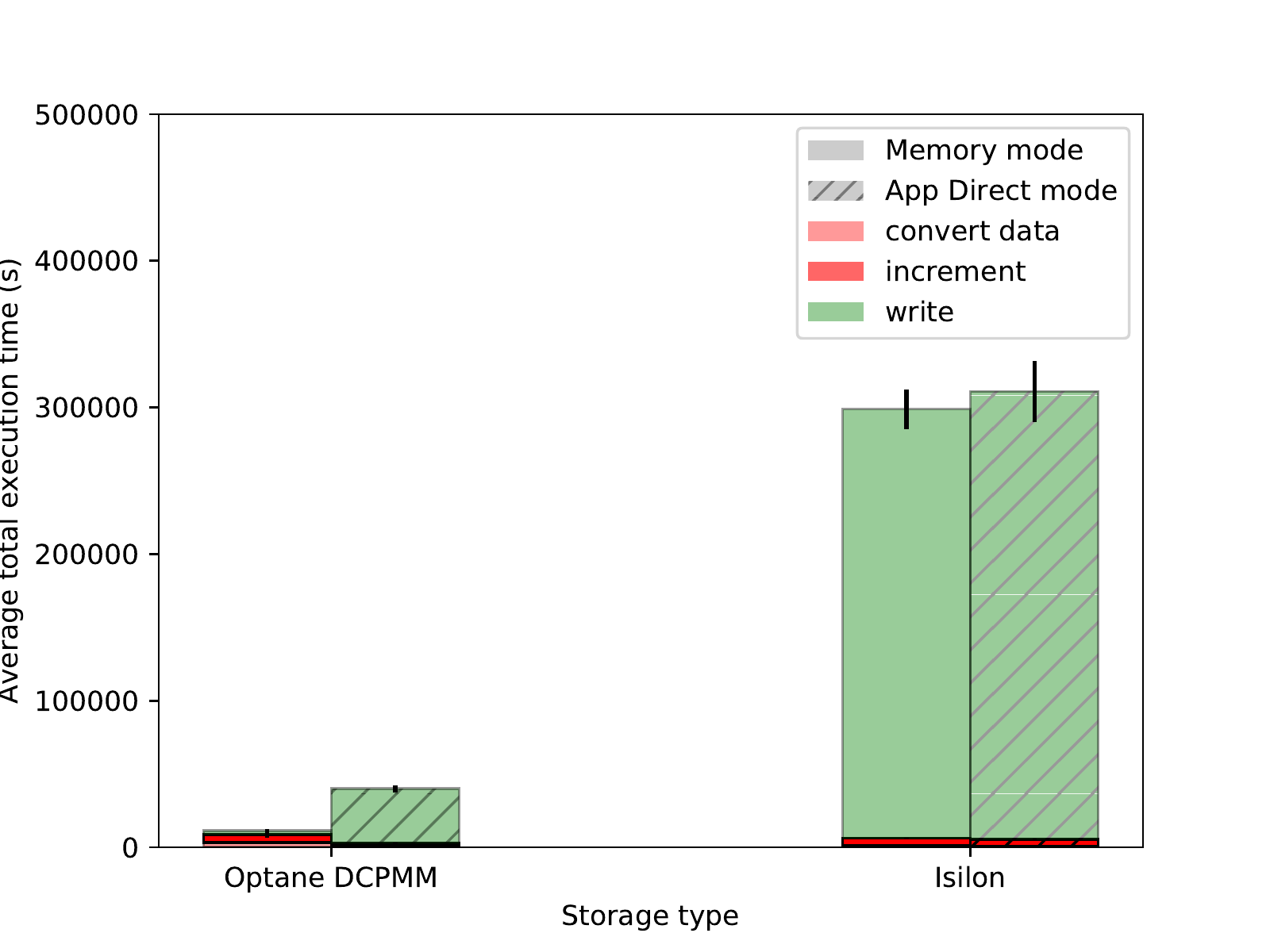}
    \caption{Total convert/increment/write breakdown}\label{fig:20totalspark96}
\end{subfigure}
\caption{Spark incrementation application processing the 20~$\mu$m BigBrain using 96 processes (three repetitions).}\label{fig:20stackedp96}
\captionsetup{belowskip=-10pt}
\end{figure*}

A notable difference in the Spark execution is that Optane DCPMM in Memory and App Direct mode
exhibit the same performance (Figure~\ref{fig:20mkspspark96}). Both of which also performed
significantly worse than the estimated I/O duration. Performance on Isilon, however, did
not really differ. The total task breakdown shows that, as expected, write times are more
significant on Isilon than on Optane DCPMM and Optane DCPMM in App Direct mode spends more time writing
than in Memory mode (Figure~\ref{fig:20totalspark96}). When comparing GNU Parallel and 
Spark implementations, Spark spent nearly half as much time writing as GNU Parallel. However,
it was found that Optane DCPMM spent more time converting the data to NiBabel than GNU Parallel spent
in loading the header.

\subsection{BIDS App Example}
\begin{figure*}
    \begin{subfigure}{\columnwidth}
        \centering
    \includegraphics[width=\columnwidth]{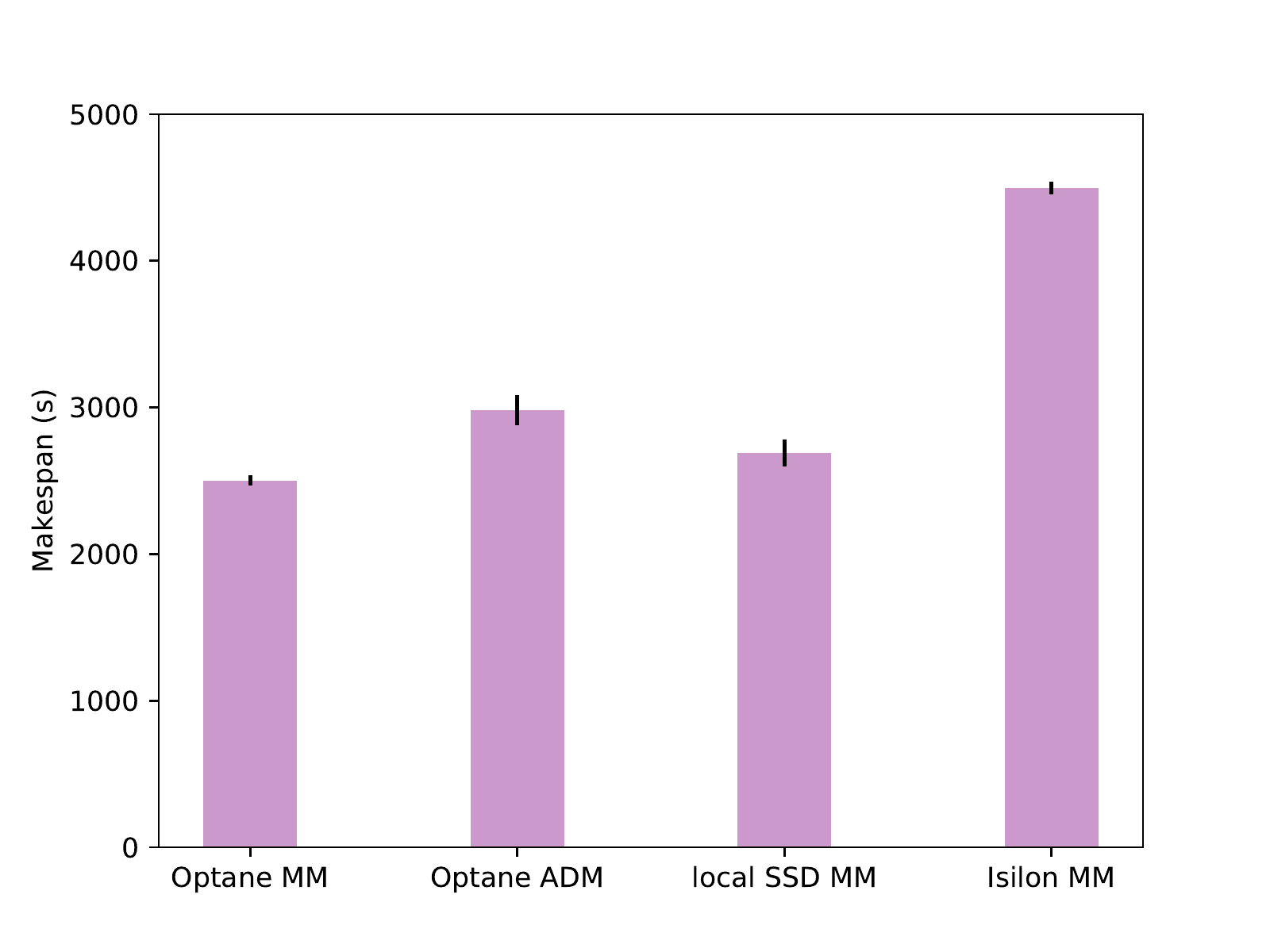}
    \caption{25 processes}\label{fig:bm25}
\end{subfigure}
    \begin{subfigure}{\columnwidth}
        \centering
    \includegraphics[width=\columnwidth]{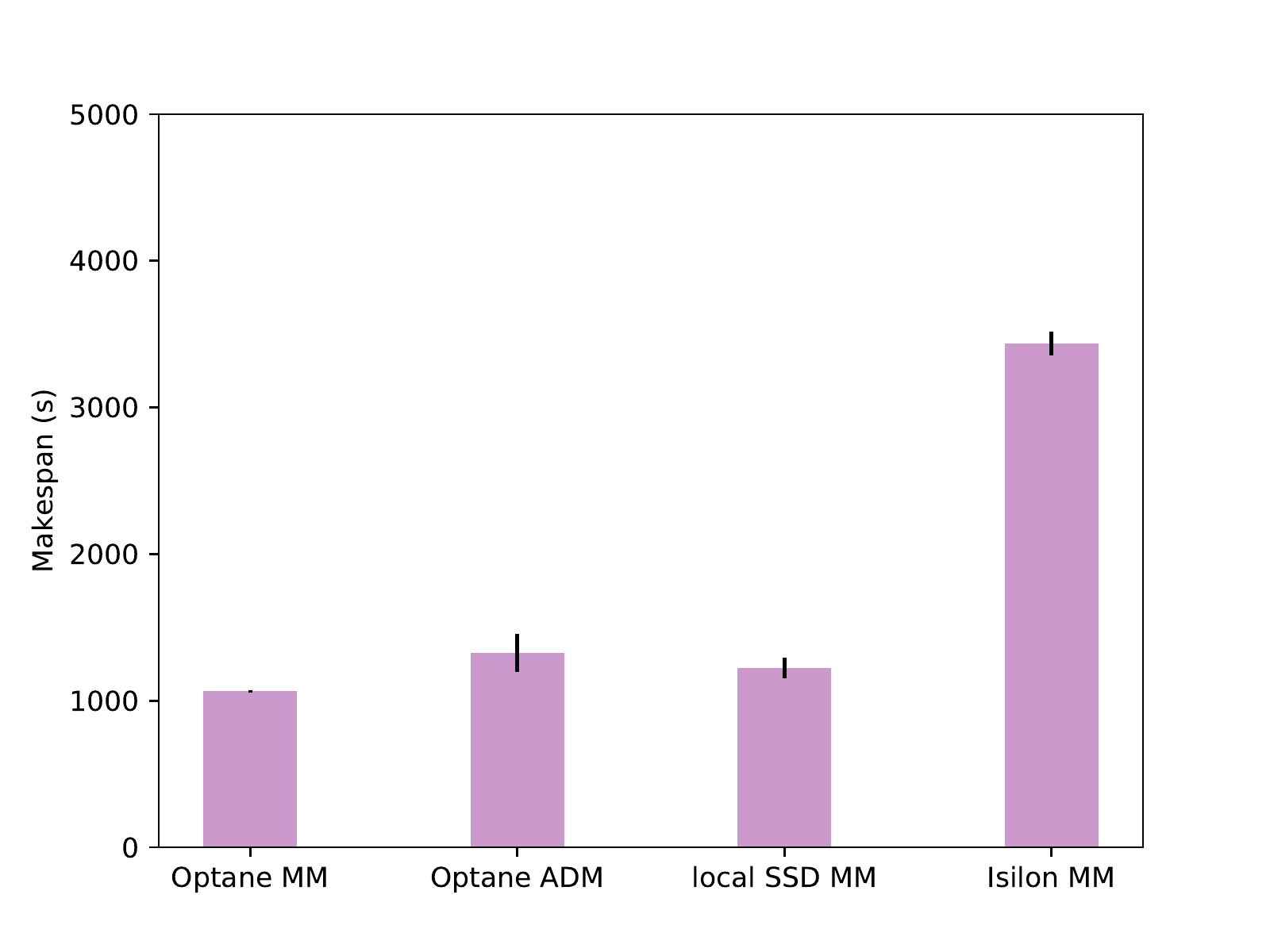}
    \caption{96 processes}\label{fig:bm96}
\end{subfigure}
\caption{Makespan of BIDS App Example 25 and 96 processes on all storage devices (three repetitions).}
\captionsetup{belowskip=-10pt}
\end{figure*}

\begin{figure*}
    \begin{subfigure}{\columnwidth}
        \centering
    \includegraphics[width=\columnwidth]{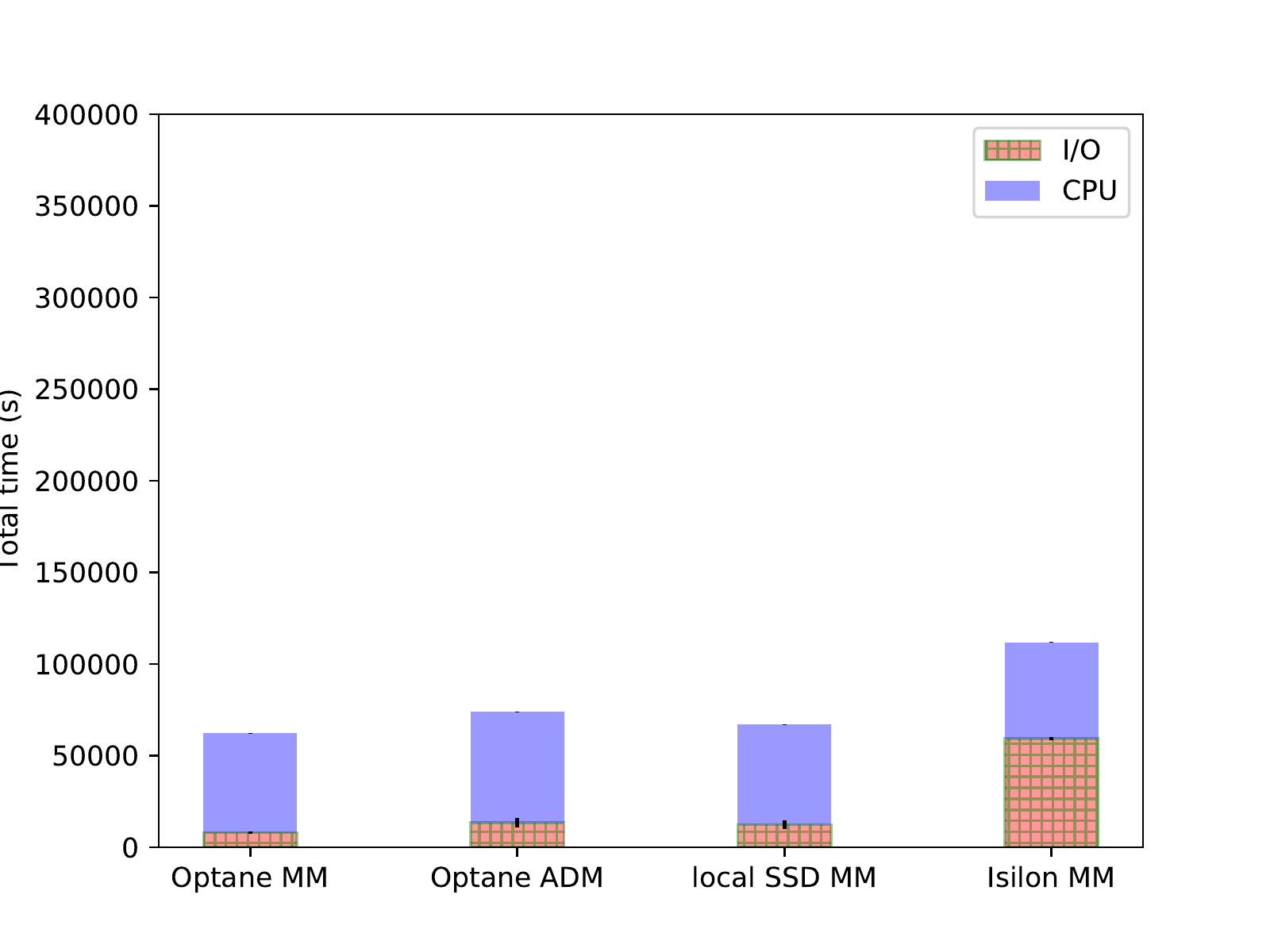}
    \caption{25 processes}\label{fig:bb25}
\end{subfigure}
    \begin{subfigure}{\columnwidth}
        \centering
    \includegraphics[width=\columnwidth]{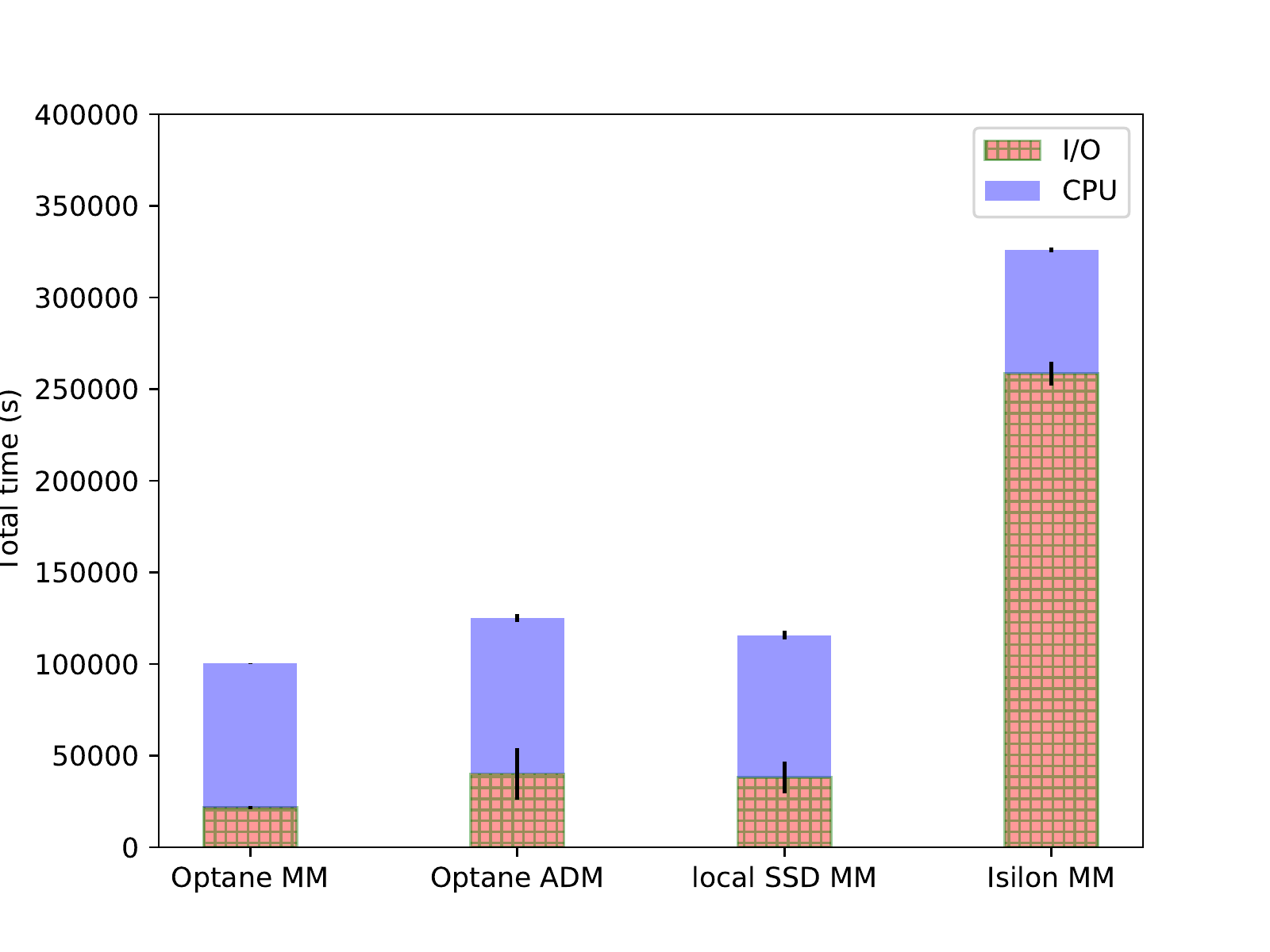}
    \caption{96 processes}\label{fig:bb96}
\end{subfigure}
\caption{I/O and CPU breakdown of BIDS App Example 25 and 96 processes on all storage devices (three repetitions).}\label{fig:bbd}
\captionsetup{belowskip=-10pt}
\end{figure*}
Unlike the BigBrain incrementation, the makespan of the BIDS App Example
executing on Optane DCPMM in App Direct mode is longer than local disk in Memory mode (Figure~\ref{fig:bm25}). However, similarly
to the BigBrain Incrementation, Optane DCPMM in Memory mode performs better than other storage devices
while Isilon has the longest makespan. The same pattern can be found with 96 process (Figure~\ref{fig:bm96}). When comparing
the executions of 25 and 96 processes, the performance improved by a similar factor on all
devices.

The CPU and I/O breakdowns (Figure~\ref{fig:bbd}) show that there is a significant increase
in the I/O time at 96 processes when compared to 25 processes. Furthermore, there appears
to also be an increase in the amount of time spent on CPU processing. In both cases,
tmpfs spends the least amount of time on I/O and CPU, followed by local disk, Optane, and Isilon.
\section{Discussion}
\subsection{Memory vs App Direct mode}

In general, the selection of storage mode did not affect overall performance. This can be
due to several reasons. For instance, the amount of available DRAM was abundant and exceeded 
all our dataset sizes. While total memory used by the application may have exceeded DRAM, as is the
case with the 20~$\mu$m \bigbrain, the amount of memory required by the application at any 
given moment did not exceed the amount of available DRAM. Although GNU Parallel would have
been able to load all the blocks in memory had we increased the block size, it would have been unlikely
that total available DRAM would have been exceeded. However, available DRAM could have been exceeded in
Memory mode when using devices that leverage memory as a write cache.
In our case, only the local SSD was set up to do so.

Optane DCPMM was the only device affected by the choice of storage mode. In all cases, with the 
exception of the Spark incrementation of the 20~$\mu$m \bigbrain, Memory mode was superior to 
App Direct mode. This is a result of Optane DCPMM using DRAM as cache in Memory mode, whereas Optane
DCPMM 
was configured as a writethrough device in App Direct mode. It is believed that had Optane DCPMM been
configured to use DRAM as a writeback cache in App Direct mode, we would have observed a similar
performance between the two. Furthermore, in the 40~$\mu$m \bigbrain executions, read time on Optane DCPMM in
App Direct mode appeared to be slower than Memory mode, potentially indicating that the Optane DCPMM was able
to keep the input data cached in DRAM, whereas in App Direct mode, the data had to be loaded from Optane DCPMM.
Therefore, it is expected the Optane DCPMM in App Direct mode may always have longer read times than Memory mode
if the data is already cached in DRAM in Memory mode.

The Apache Spark and GNU Parallel implementations differed vastly in makespan 
for Optane DCPMM in Memory mode, particularly at 96 processes. After taking a closer look at
the captured Spark metrics, what significantly varied between the Spark runs on
the different storage was garbage collection, with garbage collection taking longer on Optane DCPMM 
than on Isilon. Furthermore, garbage collection was longer in Memory mode than in
App Direct mode. It is suspected that, in these cases, task duration was so short
that garbage collection could not keep up. Moreover, Optane DCPMM in Memory mode may have longer
garbage collection if old anonymous data cannot be stored in the DRAM cache due to overall memory 
requirements, therefore requiring garbage collection to occur on slower memory.

\subsection{Effect of Page Cache}

The use of Optane DCPMM in Memory mode would enable more data to be written to available memory
rather than directly to slower storage devices. In our 40~$\mu$m \bigbrain experiments, it was found
that read times were less important on Isilon than on the local SSD (Figures~\ref{fig:stacked-25cpus}
and~\ref{fig:stacked-96cpus}). Should Isilon have benefitted from a 
writeback cache like the SSD, it is believed that performance on Isilon could have been superior 
to that of the SSD. Having Optane DCPMM as a node-local burst buffer to shared network storage 
may prove to be very beneficial in the case of Big Data neuroimaging pipelines running on
High Performance Computing clusters.

For App Direct mode applications (although also applicable to Memory mode), the 
choice of using the writeback cache depends on total amount of available memory, application
memory usage and I/O. Should the application be able to use the page cache without ever being
I/O-throttled, using a write-back cache will significantly improve performance. This can 
be seen in Figure~\ref{fig:stacked-25cpus}, for instance, where the local SSD is writing
entirely to memory, significantly reducing the total write time. However, should write operations
become throttled, this may slow down the write task duration. In such cases,
it may be preferable to set them up as write-through devices. In any case,
 despite the fact that Optane DCPMM in App Direct mode does
not leverage the writeback cache like the local SSD, is it still found to be superior in performance.
Therefore, while the performance of Optane DCPMM may not be enhanced by DRAM, it is still superior to that of
an SSD using the writeback cache and scalable network storage.

\subsection{Device scalability}
The number of storage devices attached to a particular filesystem mountpoint varied between
each storage device. For instance, DRAM consisted of 12 devices, whereas although there were 
12 Optane devices, only 6 were accessible by the mountpoint selected in App Direct mode. Furthermore,
local SSD only consisted of a single device, whereas the Isilon server was made up of 180 SSDs. The differences
in number of storage devices would account for the attainable amount of parallel writes. For instance, 
Isilon would have been the most scalable, followed by DRAM, Optane DCPMM and then the local SSD, which would
only be able to process I/O sequentially. This explains why the anticipated I/O time, is sometimes greater
than the real makespan of the pipeline.

For instance, taking into consideration the total number of DRAM devices, the expected makespan of the
40~$\mu$m \bigbrain is estimated to take around 3s, whereas sequentially it was measured to take
around 37.5s. At 25 processes, we found the DRAM makespan to be around 7s, which is much closer to the 3s
estimate than the 37.5s. Reasons for why it took longer than estimated could be that
our applications were not optimized for non-uniform memory accesses (NUMA-aware), or even that some DRAM devices were occupied, and therefore, the maximum amount of parallelism 
was not achievable. 

Despite the fact that only sequential I/O was considered for the anticipated I/O times, 
Optane DCPMM in App Direct mode always took more time than its sequential estimates. The cause
of the longer-than-expected makespan seems to be due to write times. While there may be application overheads,
they should, in theory, affect Optane DCPMM in App Direct mode as much as DRAM. While it is unknown
why Optane DCPMM writes are performing so poorly, it could be a result of filesystem configuration or 
inaccurate benchmarking of the disk. Otherwise, Optane DCPMM does not appear to scale very well with respect to
writes. Using the storage benchmarks found in Table~\ref{table:bandwidths}, it is found that with an equivalent
number of storage devices, processing the 40~$\mu$m \bigbrain using 25 processes would be faster using
local disk when compared to the real times obtained on Optane DCPMM. As there are overheads with both Optane DCPMM and local
SSD, it is uncertain that it would, in fact, be the case.

While Isilon is, in theory, the most scalable device, its performance is expected. The Isilon
server is made up entirely of Hard Disks Drives, which are the slowest type of device compared to
the others, as seen in Table~\ref{table:bandwidths}. While Isilon displays a sufficient amount of
scalability to read and write the data in a few seconds, it is a network-backed device. As a result,
performance of this device is limited by the network speed of 10Gbps. Consequently, Isilon performs at around
the same rate as sequential writes.

Another interesting aspect of Isilon is the spacing observed between the read and writes
(Figure~\ref{fig:gantt25isilon}). What occurred between the read/increment and
write operations was simply writing the previous task's benchmarks to a unique
file. This did not appear to affect any other storage device. This implies
that the latency alone of writing to Isilon was quite significant.

\subsection{Added value of persistent memory}

Our results show that Optane DCPMM has superior performance to that of other storage devices.
Optane DCPMM as a persistent memory storage device is expected to bring significant
performance improvements to the processing of neuroimaging data. Input datasets to neuroimaging pipelines
are rapidly increasing in size. Storing such datasets directly on Optane DCPMM would significantly reduce the
impacts of I/O on the processing. If Optane DCPMM is located as a persistent memory
storage device in HPC clusters,
it could also be used as a burst buffer to network-attached storage devices. This would vastly
improve the performance of standard neuroimaging pipelines, as they produce temporary data files that 
are larger than the input dataset. Writing these temporary files to slow, network-attached storage, can have
significant impacts on the performance of a pipeline.

In the case of high-resolution images, Optane DCPMM can be leveraged to enable the users to rapidly
extract their regions of interest with minimal I/O costs. Futhermore, Optane DCPMM would improve the
speed and fluidity of visualization applications on these datasets.

\subsection{Other comments}

Storage performance when scaling the application from 25 to 96 CPU threads could have been improved using a
NUMA-aware application, since memory is tied to the CPUs. Should a task requiring memory
stored on one CPU socket be sent to another CPU, this would incur some overheads, reducing overall application
and storage performance. Furthermore, Intel optimized Python and NumPy libraries could have been used to improve 
performance.

While these experiments may not accurately reflect the maximum performance of the storage devices for the applications,
they are meant to reflect the performance that would be obtained by the average neuroimaging researcher
executing standardized tools on infrastructure made available to them.

\section{Conclusion}

Optane DCPMM has been found to drastically reduce the processing time of neuroimaging applications and
can bring the performance close to DRAM speeds. Extending available memory using Optane DCPMM is also 
expected to reduce I/O times of write-back devices by extending available cache space.
It is unclear, however, if SSDs can perform just as well as Optane DCPMM with additional storage devices
attached.

It is believed that Optane DCPMM can be useful for a variety of neuroimaging applications. For
instance, typically applications executing in HPC environments can benefit from 
the speedups provided by Optane DCPMM. Furthermore, image visualization servers will be able
to display higher resolution images at significantly greater speeds.
\section{Acknowledgement}
We are very grateful to the Intel and Dell-EMC teams for giving us the
opportunity to experiment with the Intel Optane DC Server, for providing
high-quality technical support, and for their thorough feedback on the
manuscript. 
\bibliographystyle{IEEEtran} 
\bibliography{biblio}

\end{document}